\documentclass[12pt,aps,pra,floatfix]{revtex4-1}
\usepackage{graphicx}% Include figure files
\usepackage{geometry}
\usepackage{bm}% bold math
\usepackage{amsmath}
\usepackage{subfigure}
\usepackage{amssymb}%
\usepackage{tabularx} % in the preamble
\usepackage{color} 
\usepackage{ulem}
\usepackage[mathlines]{lineno}% Enable numbering of text and display math
\begin{document}

%\preprint{}

\title{Longitudinal beam dynamics with active cavity systems}% Force line breaks with \\
\author{Chao Li}
\altaffiliation[]{li.chao@desy.de}
\author{Yong-Chul Chae}
\altaffiliation[]{yong-chul.chae@desy.de}
\author{Sven Pfeiffer}
\author{ Chris Christou}
\affiliation{Deutsches Elektronen Synchrotron, Notkestrasse 85, 22607 Hamburg, Germany}
\date{\today}

\begin{abstract}
In storage-ring-based light sources, harmonic cavities are commonly employed to lengthen the bunch, thereby mitigating collective effects and enhancing beam lifetime. However, this dual-RF configuration also introduces significant challenges, as the impedance of the fundamental modes can drive the beam into a longitudinal coupled-bunch instability regime. To mitigate this effect, low-level radio-frequency (LLRF) feedback is proposed to reduce the effective impedance experienced by the beam. This work investigates longitudinal beam dynamics in the PETRA-IV dual-RF system with normal-conducting cavities, explicitly accounting for the LLRF feedback loop. Both analytical modeling and numerical simulations are employed to characterize the onset and growth of coupled-bunch instabilities. The results show that, with appropriately defined LLRF specifications, the destabilizing influence of the cavity fundamental modes can be effectively suppressed, enabling stable operation of the storage ring at the design current. This work highlights the critical role of RF feedback systems in ensuring robust longitudinal stability, thereby supporting the realization of PETRA-IV's design goals and contributing to the broader development of next-generation synchrotron light sources, where high brilliance and operational reliability are essential.

\end{abstract}
\maketitle
%\tableofcontents  % <-- generates the content page
%\newpage          % start main text on a new page

\section{Introduction}\label{section:1}
In high-intensity circular accelerators, various longitudinal beam instabilities can be caused by interactions between the circulating beam and the impedance modes of the accelerating cavities \cite{AlexChao,ng2006physics}. If the higher-order modes (HOMs) are well damped in the cavity design, the fundamental mode becomes the main concern. In addition to interacting with the beam, the cavity fundamental mode is also coupled to at least two other components. One is the generator current, which excites the cavity field that acts on the beam. The other is the LLRF feedback control loop, which regulates the cavity field experienced by the beam. To obtain a reliable assessment of beam performance, these coupled dynamics must be studied consistently~\cite{berenc2015modeling}.
   
In synchrotron light sources, many facilities are being upgraded to the fourth generation, characterized by natural emittances at the $10^{-11}$ m level, bunch lengths at the $10^{-3}$ m level, and beam currents at the $10^{-1}$ A level~\cite{agapov2024beam,PhysRevSTAB.12.120701,PhysRevAccelBeams.24.110701,sajaev2025commissioning,boge2025sls}. One of the major concerns is maintaining a reasonable Touschek lifetime. To ensure sufficient Touschek lifetime under different beam conditions, additional active or passive, normal-conducting or superconducting harmonic RF cavities can be introduced to lengthen the bunch and reduce the electron density, thereby mitigating the Touschek effect. The harmonic cavities also introduce a significant synchrotron tune spread, which can stabilize collective beam motion through the Landau damping mechanism~\cite{landau196561,chin1982instability,herr2016introduction,PhysRevAccelBeams.24.011002,PhysRevAccelBeams.23.074402}. However, the fundamental modes of the harmonic cavities are also significant sources of impedance and may, conversely, increase the chance of exciting collective instabilities~\cite{PhysRevAccelBeams.21.114404,lindberg2018theory}.

PETRA-III is going to be upgraded to PETRA-IV, which will be a fourth-generation storage-ring-based light source capable of delivering a beam with an emittance as low as 20~pm. Tab.~\ref{tab:1} gives the nominal design parameters of the storage ring. Two operation schemes are proposed for the PETRA-IV storage ring: 1) timing mode and 2) brightness mode. In the timing mode, 80 bunches are evenly distributed around the ring, with a total beam current of 80~mA. In the brightness mode, 1920 bunches are evenly distributed around the ring, with a total beam current of 200~mA. For PETRA-IV, a third-harmonic normal-conducting RF EU-cavity~\cite{PEREZ2025170195}, designed by ALBA, has been chosen as the baseline. Tab.~\ref{tab:2} lists the fundamental-mode RF parameters of the main and third-harmonic cavities.
\begin{table*}[!htp]
\caption{Nominal lattice parameters of PETRA-IV H6BA lattice.}
\centering
\begin{tabular}{|l|r|r|r|r|}\hline \hline
\textbf{Parameters}   &\textbf{units}    &\textbf{symbol}       &  \textbf{DW Closed}  & \textbf{DW Open} \\ \hline
Energy      &GeV       &$E_0$          & 6           & 6     \\ \hline
Circumference  &m      &$C$          & 2304        & 2304  \\ \hline
Natural Emittance  &pm          &$\epsilon_0$   & 20          & 43    \\ \hline
Emittance Ratio &          &$\kappa$         & 0.1         & 0.1   \\ \hline
Tunes           &          &$\nu_x$/$\nu_y$  & 135.18/86.27  & 135.18/86.23   \\ \hline
Momentum Compact Factor &         &$\alpha_c$            & 3.33$\times 10^{-5}$    & 3.33$\times 10^{-5}$   \\ \hline
Damping Time  &ms        &$\tau_x$ & 17.76      &  39.23   \\ \hline
Damping Time  &ms        &$\tau_y$ & 22.14      & 69.63    \\ \hline
Damping Time  &ms        &$\tau_s$  & 12.62      & 56.84    \\ \hline
Natural Energy Spread &rad        &$\sigma_e$    &  8.9$\times 10^{-4}$    & 7.37$\times 10^{-4}$  \\ \hline
Natural Bunch Length  &mm     &$\sigma_s$ & 2.3        & 1.794  \\ \hline
Energy Loss    &MeV        & $U_0$  & 4.2      & 1.423  \\ \hline
Main Cavity Voltage &MV        &$V_{c,1}$ & 8          & 8      \\ \hline
Main Cavity Harmonics &      &$h$        & 3840       & 3840    \\ \hline
\multicolumn{5}{|l|}{\textbf{DW: Damping Wiggler}} \\  \hline 
\end{tabular}
\label{tab:1}
\end{table*}

\begin{table}[htp]   
\caption{RF parameters of PETRA-IV storage ring, 24 identical cavities for each type.}
\centering
\begin{tabular}{|l|r|r|r|}\hline \hline
\textbf{Parameter}  &\textbf{Symbol}  &\textbf{Main RF ($n=1$)}    &\textbf{Harmonic RF ($n=3$)} \\ \hline
RF Freq. (Hz)   & $f_{rf,n}$   & 4.996$\times 10^8$  & 1.499$\times 10^9$  \\  \hline
Coupling Factor & $\beta_n$    & 5        & 5    \\ \hline
Shunt Impedance ($\Omega$)   & $R_{s,n}$   & 81.6$\times 10^6$    & 36.0$\times 10^6$    \\ \hline
%Half Bandwidth (KHz) (measured)  & $f_{1/2}$ &  24.6  & 277  \\
Half Bandwidth (Hz)   & $f_{1/2}$ &  50.6$\times 10^3$  & 264.5$\times 10^3$ \\ \hline
%Quality Factor( design / measured) & $Q_{0,n}$   & 29600 /40623   & 17000 / 17046  \\
Quality Factor (design) & $Q_{0,n}$   & 29600   & 17000   \\ \hline
Detuned Frequency (Hz)  & $df_{rf,n}$  & -8.928$\times 10^3$  & 46.64$\times 10^3$  \\ \hline
\end{tabular}
\label{tab:2}
\end{table} 

In this paper, we discuss in detail the longitudinal beam dynamics of the dual-RF system in the PETRA-IV storage ring. In Sec.~\ref{section:2}, we introduce several topics. Although most of these topics have already been discussed by other researchers~\cite{PhysRevSTAB.4.074401,alves2025theoretical,PhysRevAccelBeams.28.054401,PhysRevAccelBeams.20.011004,PhysRevSTAB.13.052802}, we still believe that it is helpful to present them together in a single paper to clarify the mechanisms of different instabilities. In Sec.~\ref{section:2.A}, we show how to model the interaction between the generator current and the cavity fundamental modes~\cite{PhysRevAccelBeams.26.114602}. In Sec.~\ref{section:2.B}, we discuss the static component of the beam-induced voltage. In Sec.~\ref{section:2.C}, we discuss the longitudinal particle dynamics in general. The methods for calculating the longitudinal oscillation frequency and its spread under equilibrium beam conditions are also presented. In Sec.~\ref{section:2.H}, we present the DC Robinson instability criterion, which is closely related to the static beam-induced voltage. The dynamic component of the beam-induced voltage and the corresponding instabilities are discussed in Secs.~\ref{section:2.D}--\ref{section:2.G}. The numerical method used to simulate the beam-induced voltage is presented in Sec.~\ref{section:2.I}. In Sec.~\ref{section:3}, we evaluate the longitudinal instabilities in the PETRA-IV storage ring. It is found that the coupled-bunch instability driven by the third-harmonic cavity is the main limitation to reaching the design beam current. To mitigate this instability, we propose the use of LLRF feedback to reduce the effective impedance experienced by the beam. The model of the LLRF feedback loop and its application to PETRA-IV are discussed in Secs.~\ref{section:4} and \ref{section:5}, respectively. Finally, the summary and outlook are presented in Sec.~\ref{section:6}.

\section{Longitudinal dynamics}\label{section:2}
\begin{figure}[!htp]
\centering
\includegraphics[width=1\linewidth]{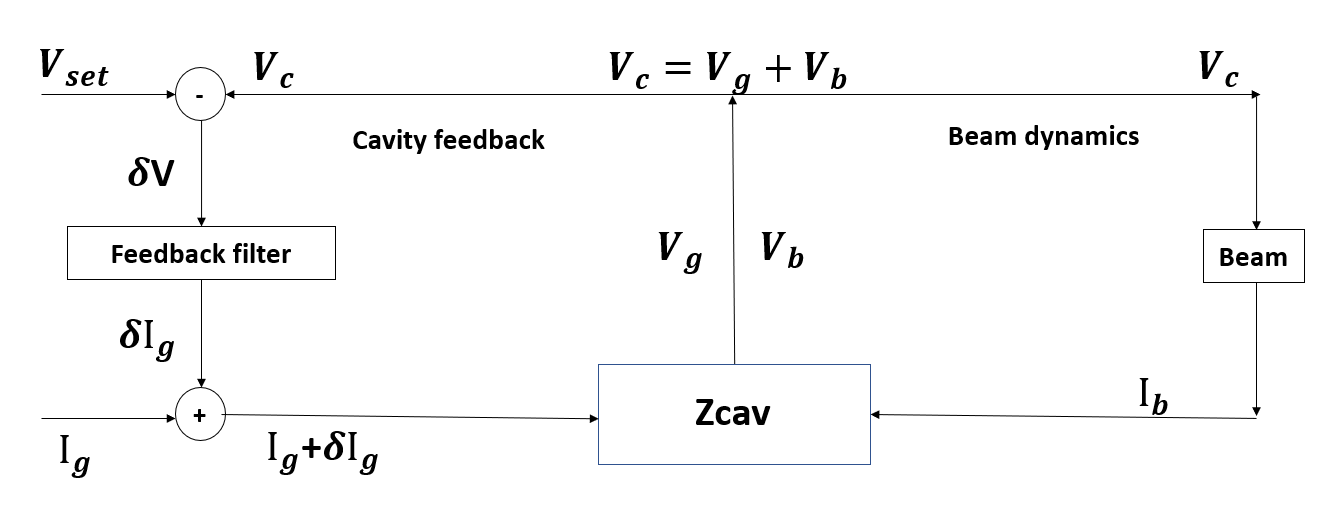} 
\caption{Interaction between generator dynamics, beam dynamics, and cavity feedback.}
\label{fig:beam_cavity_feedback}
\end{figure}

The impedance of the cavity fundamental mode $\boldsymbol{Z}(\omega)$ is usually modeled as an RLC circuit. 
\begin{equation}
\label{eq:2.1}
\boldsymbol{Z}(\omega) = \frac{R_L}{1+i Q_L(\frac{\omega_r}{\omega}-\frac{\omega}{\omega_r})} = R_L \cos\psi e^{-i\psi} 
\end{equation}
where $\omega_r$, $Q_L$ and $R_L$ are the angular resonant frequency, the loaded quality factor and shunt impedance respectively, $i$ is the imaginary unit;  $\psi$ is the detuning angle defined as 
\begin{equation}
\label{eq:2.2}
\tan \psi = Q_L(\frac{\omega_r}{\omega}-\frac{\omega}{\omega_r}),
\end{equation}
Fig.~\ref{fig:beam_cavity_feedback} shows the sketch of the dynamics in an active RF cavity considering the fundamental mode only.  In the time domain, the voltage experienced by the beam is given by  
\begin{equation}
\label{eq:2.3}
\boldsymbol{V_c}(t) = \boldsymbol{V_{g}}(t) + \boldsymbol{V_{b}}(t) 
\end{equation}
where $\boldsymbol{V_{g}}(t)$  and $\boldsymbol{V_{b}}(t)$ are the voltages excited by the generator current $\boldsymbol{I_{g}}(t)$ and beam current $\boldsymbol{I_{b}}(t)$ respectively. In the following sections, the dynamics driven by the generator current $\boldsymbol{I_{g}}(t)$ and the beam current $\boldsymbol{I_{b}}(t)$, as well as the particle motion and collective instabilities, will be discussed.

\subsection{Generator dynamics}\label{section:2.A}
The generator current is a continuous-wave (CW) signal with a single frequency component $\omega_{g}=\omega_{rf}$, so that $\boldsymbol{I_{g}}(t)=\boldsymbol{\tilde{I}_{g}}(t)e^{i\omega_{g}t}$. Together with the assumed RLC impedance model, the excited  generator voltage $\boldsymbol{\tilde{V}_g}(t)$ follows~\cite{schilcher1998vector} the differential equation 
\begin{equation}
\label{eq:2.A.1}
\frac{d^2}{dt^2}\boldsymbol{\tilde{V}_g}(t) + \frac{\omega_{r}}{Q_{L}} \frac{d}{dt}\boldsymbol{\tilde{V}_g}(t) + \omega_{r}^2 \boldsymbol{\tilde{V}_g}(t) = \frac{\omega_{r} R_{L}}{Q_L} \frac{d}{dt}\boldsymbol{\tilde{I}_g}(t),
\end{equation}
For a given initial driving current $\boldsymbol{\tilde{I}_{g}}(t=0)$, Eq.~\ref{eq:2.A.1} can be numerically solved in the time domain.

If the following conditions are met: (1) the second-order terms can be neglected; (2) $Q_L\gg1$; (3) $\Delta\omega=\omega_r-\omega_{g}\ll\omega_r$, then together with the zero-order hold method~\cite{schilcher1998vector,berenc2015modeling}, the solution of Eq.~\ref{eq:2.A.1} can be further simplified in the state space    
\begin{equation}
\label{eq:2.A.2}
\begin{aligned}
\left(\begin{array}{c} 
\Re \boldsymbol{\tilde{V}}  \\ 
\Im \boldsymbol{\tilde{V}}   
\end{array}\right)_{t+\Delta t}
&= e^{-\omega_{1/2} \Delta t}
\left(\begin{array}{cc} 
\cos \Delta \omega \Delta t  & -\sin \Delta \omega \Delta t     \\ 
\sin \Delta \omega \Delta t  &  \cos \Delta \omega \Delta t 
\end{array}\right)
\left(\begin{array}{c}
\Re \boldsymbol{\tilde{V}}  \\ 
\Im \boldsymbol{\tilde{V}}  
\end{array}\right)  \\
 &+ \frac{\omega_r R_L}{2 Q_L (\omega_{1/2}^2 + \Delta \omega^2) }
\left(\begin{array}{cc}
A  & B   \\ 
-B   &  A
\end{array}\right)
\left(\begin{array}{c}
\Re \boldsymbol{\tilde{I}}  \\ 
\Im \boldsymbol{\tilde{I}} 
\end{array}\right)_t \\
\end{aligned}
\end{equation}
where  $\Re$ and $\Im$ denote the real and imaginary parts of the phasor, $\omega_{1/2}=1/\tau_f=\omega_r/(2Q_L)$ is the cavity half-bandwidth,  $\Delta t$ is the time step, $A$ and $B$ are
\begin{equation}
\label{eq:2.A.3}
\begin{aligned}
A &= \Delta \omega e^{-\omega_{1/2} \Delta t} \sin \omega_{1/2} \Delta t - \omega_{1/2} e^{-\omega_{1/2} \Delta t} \cos \omega_{1/2} \Delta t + \omega_{1/2}  \\
B &= \omega_{1/2}  e^{-\omega_{1/2} \Delta t} \sin \omega_{1/2} \Delta t + \Delta \omega e^{-\omega_{1/2} \Delta t} \cos \omega_{1/2} \Delta t - \Delta\omega.
\end{aligned}
\end{equation}
Eq.~\ref{eq:2.A.3} can be implemented in numerical simulations to model the generator dynamics \cite{elegantWebPage,berenc2015modeling,LI2025170031,CETASim}. 

\subsection{Beam-induced voltage: the static term}\label{section:2.B}
Unlike the generator current $\boldsymbol{I_g}(t)$, the beam current (bunch train) $\boldsymbol{I_b}(t)$ is discrete. If we assume that $M$ bunches uniformly occupy the ring, staying at the equilibrium position (or all bunches have the same offset $\tau_0$) without oscillation, and denote the normalized distribution of each bunch as $\boldsymbol{\rho}(t)$, then the beam current signal in the time domain is 
\begin{equation}
\label{eq:2.B.1}
\boldsymbol{I}_b(t) = - e N \sum_{k=-\infty}^{k=\infty} \boldsymbol{\rho}(t) \delta[t-\frac{k T_0}{M} - \tau_0],  
\end{equation}
where $e$ is the electron charge, $N$ is the number of particles per bunch and $T_0$ is the revolution time. Eq.~\ref{eq:2.B.1} shows that the beam current phasor has only a real component and always points to the negative axis. Correspondingly, the beam spectrum can be written as  
\begin{equation}
\label{eq:2.B.2}
\begin{aligned}
\boldsymbol{I}_b(\omega) &= - e N M \omega_0  \sum_{p=-\infty}^{p=\infty} \boldsymbol{\rho}(\omega) \delta[\omega-p M \omega_0] e^{i \omega \tau_0},
\end{aligned}
\end{equation}
where $\boldsymbol{\rho}(\omega)$ is the spectrum of $\boldsymbol{\rho}(t)$. Together with the cavity impedance $\boldsymbol{Z}(\omega)$, the beam-induced voltage in the frequency and time domains can be obtained as 
\begin{equation}
\label{eq:2.B.3}
\begin{aligned}
\boldsymbol{V_b}(\omega) &=\boldsymbol{I_b}(\omega) \boldsymbol{Z}(\omega) =- e N M \omega_0 \sum_{p=-\infty}^{p=\infty} \boldsymbol{\rho}(\omega')  \boldsymbol{Z}( \omega') e^{i \omega' \tau_0}   \\
\boldsymbol{V_b(t)}&=\int \frac{d \omega}{2 \pi}e^{-i \omega t} \boldsymbol{V}_b(\omega) =  - \frac{N M e}{T_0}  \sum_{p=-\infty}^{p=\infty} \boldsymbol{\rho}(\omega')  \boldsymbol{Z}( \omega') e^{-i \omega' (t-\tau_0)},
\end{aligned}
\end{equation}
where $\omega'=pM\omega_0$. Noticeably, the frequency component at $\omega'=\pm\omega_{g}=\pm h \omega_0$ is synchronized with the bunch repetition rate and $\boldsymbol{Z}( \omega)$ is narrow around $\omega_g$, then Eq.~\ref{eq:2.B.3} can be further simplified to  
\begin{equation}
\label{eq:2.B.4}
\begin{aligned}
\boldsymbol{V_b}(t) %&=  - I_{0}  (\boldsymbol{\rho}(-\omega_{g}) \boldsymbol{Z}(-\omega_{g})e^{i\omega_g t}+\boldsymbol{\rho}(\omega_{g}) \boldsymbol{Z}(\omega_{g})e^{-i\omega_g t}) \\
%&=- I_{0} \boldsymbol{\rho}(\omega_{g}) R_L \cos\psi(e^{-i\psi} e^{i \omega_{g} t} +  e^{i\psi} e^{-i \omega_{g} t}) \\
&=- 2I_{0} \boldsymbol{\rho}(\omega_{g}) R_L \cos\psi e^{i\psi} e^{i \omega_{g} (t-\tau_0)},
\end{aligned}
\end{equation}
where $I_{0} = N M e / T_0$ is the DC  beam current, $\psi$ is the detuning angle at $\omega=\omega_g$. The term $ \boldsymbol{\rho}(\omega_{g})$ is interpreted as the bunch form factor. If the following bunch has the same shape as the preceding bunch, the beam-induced voltage sampled by the following bunch can be obtained by multiplying $\boldsymbol{V_b(t)}$ by another form factor $\boldsymbol{\rho}(\omega_{g})$, 
\begin{equation}
\label{eq:2.B.5}
\begin{aligned}
\boldsymbol{V_b}(t=k/MT_0 + \tau_0) = -V_{br} \boldsymbol{\rho^2}(\omega_{g}) \cos\psi e^{i\psi},
\end{aligned}
\end{equation}
where $V_{br} = 2 I_0 R_L$. It shows that, when all bunches share the same offset $\tau_0$, the beam-induced voltage provides a constant kick to all bunches. It reflects the fact that the beam-induced voltage moves synchronously with coherent beam motion. Hereafter, we  set $\tau_0=0$ at the bunch center by default. This static beam-induced voltage given by  Eq.~\ref{eq:2.B.4}, oscillates similarly to an ideal RF element and does not lead to any collective instability. The solution given by Eq.~\ref{eq:2.B.4} is usually applied to set up the generator voltage $\boldsymbol{V_g} = \boldsymbol{V_{c}} - \boldsymbol{V_{b}}$ at the initial time with $t=0$, $\tau_0=0$. Dropping the term $e^{i \omega_{g} t}$, the relation in the frame rotating at $\omega_{g}$ is simplified to $\boldsymbol{\tilde{V}_g} = \boldsymbol{\tilde{V}_{c}} - \boldsymbol{\tilde{V}_{b}}$. 

If the broadband impedance is introduced, the beam profile $\boldsymbol{\rho}(t)$ generally becomes asymmetric. In this scenario, the beam spectrum $\boldsymbol{\rho}(\omega)$ is a complex function \cite{PhysRevSTAB.17.064401} so that the corresponding factor at $\omega_g$ is $\boldsymbol{\rho^2}(\omega_{g})e^{i\phi_{\rho}}$ in Eq.~\ref{eq:2.B.5}. By modifying the generator voltage  $\boldsymbol{\tilde{V}_g}$, the required $\boldsymbol{\tilde{V}_c}$ can be reached accordingly.  

\subsection{Dynamics of beam longitudinal motion}\label{section:2.C} 
In this subsection, we discuss the longitudinal beam dynamics within a single bunch when the beam-induced voltage includes only the static term. The motion of the particles is described in the frame rotating at $e^{i\omega_{g}t}$. Denote $p_0$ and $E_0$ as nominal momentum and energy of the synchrotron particle; choose $t=0$ as the synchronous time which covers the energy loss $U_0$ from synchrotron radiation damping, $\Re[ \boldsymbol{\tilde{V}_c}]=U_0/e$; denote the non-synchrotron particle with momentum $p$ and energy $E$ and  momentum offset $\delta_p=(p-p_0)/p_0$.  If the beam-induced voltage can be ignored, the voltage experienced by a particle with a time offset $\tau$ is $\Re[\boldsymbol{\tilde{V_c}} e^{i \omega{g} \tau}]=\Re[\boldsymbol{\tilde{V_g}} e^{i \omega_{g} \tau}]$. Normally, the synchrotron oscillation amplitude in time, $\tau$, is small in electron storage rings. Therefore, using a Taylor expansion,  
\begin{equation}
\label{eq:2.C.1}
\boldsymbol{\tilde{V}_c} e^{i \omega_{g} \tau} = \boldsymbol{\tilde{V}_c} + \sum_{k=1} \boldsymbol{\tilde{V}_c^k} \frac{\tau^k}{k!}, 
\end{equation}
where $\boldsymbol{\tilde{V}_c^{k}}$ represents the value of the $k$th derivative of $\boldsymbol{\tilde{V}_c}$ with respect to $\tau$ at $\tau=0$. Using Eq.~\ref{eq:2.C.1},  $\delta_p$ evolves according to
\begin{equation}
\label{eq:2.C.2}
\frac{\delta_p}{T_0} = \dot{\delta_p} = \frac{e}{T_0E_0 \beta^2}\sum_{k=1} \Re [\boldsymbol{\tilde{V}_c^{k}}] \frac{\tau^k}{k!}.
\end{equation}
Meanwhile, the time offset $\tau$ evolves according to   
\begin{equation}
\label{eq:2.C.3}
\dot{\tau} = \eta \delta_p  \quad
\end{equation}
Combining Eq.~\ref{eq:2.C.2} and Eq.~\ref{eq:2.C.3}, the complete longitudinal particle dynamics are obtained as
\begin{equation}
\label{eq:2.C.4}
\ddot{\tau} = \frac{e\eta }{T_0E_0 \beta^2}\sum_{k=1} \boldsymbol{\Re \tilde{V}_c^{k}} \frac{\tau^k}{k!} = -\frac{dU_c(\tau)}{d\tau} 
\end{equation}
where $U_c$ is the equivalent potential term in the Hamiltonian, as given in Eq.~\ref{eq:2.C.13}, and can be explicitly expressed as
\begin{equation}
\label{eq:2.C.5}
\begin{aligned}
U_c(\tau) & =-\frac{e\eta }{T_0E_0 \beta^2}  \Re[ \boldsymbol{ \tilde{V}_c} ( \frac{e^{i\omega_g \tau}-1}{i\omega_g} -\tau)]  \\
&= -\frac{e\eta }{T_0E_0 \beta^2} \sum_{k=1} \boldsymbol{\Re \tilde{V}_c^{k}} \frac{\tau^{k+1}}{(k+1)!} = \sum_{k=1} a_k \tau^{k+1}.
\end{aligned}
\end{equation}
The coefficient $a_k$ is given by
\begin{equation}
\label{eq:2.C.6}
\begin{aligned}
a_k &=- \frac{e\eta }{T_0E_0 \beta^2}  \boldsymbol{\Re}[ \frac{(i\omega_g)^k}{(k+1)!}  \boldsymbol{\tilde{V}_c}]. 
\end{aligned}
\end{equation}
The synchrotron frequency $\omega_s$ can be defined by retaining only the $a_1$ term:
\begin{equation}
\label{eq:2.C.7}
\omega_s^2 = 2 a_1. 
\end{equation}

In Eq.~\ref{eq:2.C.5}, we define the potential term $U_c(\tau)$, including the static beam-induced voltage. If the broadband ring impedance $\boldsymbol{Z_B}(\omega)$ is also taken into account, we can define a general equivalent potential well as $U(\tau)=U_c(\tau)+ U_B(\tau)$, where
\begin{equation}
\label{eq:2.C.10}
U_B(\tau) =- \frac{\eta }{T_0E_0 \beta^2} eN\omega_0 \int_0^{\tau} d\tau \int \frac{d \omega}{2 \pi}e^{-i \omega t} \boldsymbol{\rho}(\omega) \boldsymbol{Z_B}(\omega),
\end{equation}
Finally, the total Hamiltonian of a particle is given by
\begin{equation}
\label{eq:2.C.13}
H(\delta_p,\tau) = \frac{\eta}{2} \delta_p^2 + \frac{U(\tau)}{\eta}
\end{equation}
from which the beam profile can be obtained by numerically solving the Haissinski equation~\cite{Haissinski}:
\begin{equation}
\label{eq:2.C.11}
\rho(\tau) = \rho_0 \exp(-\frac{U(\tau)}{ \eta^2 \sigma_{\delta}^2}),
\end{equation}
where $\sigma_{\delta}$ is the rms energy spread, $\rho_0$ is the normalized factor. Once the profile $\rho(\tau)$ is obtained, the average and rms values of any function $f(\tau)$ can be calculated accordingly:
\begin{equation}
\label{eq:2.C.12}
\bar{f}(\tau) = \frac{\int f(\tau) \rho(\tau) d \tau}{\int \rho(\tau) d \tau}, \quad \quad \Delta f(\tau) = \sqrt{\frac{\int (f(\tau) -\bar{f}(\tau))^2 \rho(\tau) d \tau}{\int \rho(\tau) d \tau}}.
\end{equation}

Two particularly important quantities are the average synchrotron frequency $\bar{\omega_s}$ and its spread $\Delta\omega_s$, since they are closely related to the beam spectrum sidebands and Landau damping. Here, we introduce a simple numerical method to calculate these values. By rearranging Eq.~\ref{eq:2.C.3}, the total time required for a particle with a given $H$ to complete one closed orbit in the longitudinal phase space can be obtained as
\begin{equation}
\label{eq:2.C.14}
\Delta t = 2 \int_{\tau_{-}}^{\tau_{+}} \frac{d\tau}{\eta \delta_p} = 2 \int_{\tau_{-}}^{\tau_{+}} \frac{d\tau}{\sqrt{2 (\eta  H-U(\tau))}},
\end{equation}
where the integration limits correspond to the two turning points in the longitudinal phase space, at which $\dot{\tau}=\eta \delta_p = 0$. Finally, the synchrotron tune of the particle on this Hamiltonian torus can be obtained as $\nu_s(H) = T_0/\Delta t$.

\if(false)
Assume that the beam profile follows the Gaussian profile in $\delta_p$ space, the longitudinal tune can be found by one more weighing from energy spread point of view. 
\begin{equation}
\label{eq:2.C.15}
\bar{\nu_s} =  \frac{\int \nu_s(\delta_{p.m}) \rho_p(\delta_{p,m}) d \delta_p}{\int \rho_p d \delta_p} ,
\end{equation}

\begin{equation}
\label{eq:2.C.16}
 dt = \frac{d\tau}{\eta \delta_p}.
\end{equation}
%To this value, we have to switch to the action-angle frame defined as  
\begin{equation}
\label{eq:2.C.17}
J(H) = \frac{1}{2 \pi} \oint \delta d\tau=  \frac{1}{2 \pi} \int_{\tau_{-}}^{\tau_{+}} \sqrt{2/\eta(H-U(\tau))} d\tau
\end{equation}
where the integration is along the trajectory in the longitudinal phase space $(\tau,\delta)$.  
\fi

Noticeably, the equation of the particle motion discussed in this subsection is in general. It can be expanded to multi-RF system by setting $\boldsymbol{\tilde{V}_{c}} = \sum_n \boldsymbol{\tilde{V}_{c,n}}$, where $n$ is the cavity index.

\subsection{DC Robinson instability}\label{section:2.H}
As discussed in Sec.~\ref{section:2.B}, when all bunches share the same offset $\tau_0$ (corresponding to the $\mu=0$ coupled-bunch-mode (CBM) discussed in Sec.~\ref{section:2.E}), the static beam-induced voltage is constant as shown by Eq.~\ref{eq:2.B.5}. In this case, the voltage experienced by a particle is $\Re[\boldsymbol{\tilde{V}_c} e^{i \omega_{g} \tau_0}]=\Re[\boldsymbol{\tilde{V}_g}e^{i \omega_{g} \tau_0} + \boldsymbol{\tilde{V}_b}]$.  
Starting from this equation, all the derivations from Eq.~\ref{eq:2.C.1} to Eq.~\ref{eq:2.C.7} are still available by replacing $\boldsymbol{\tilde{V}_c^k}$ with $\boldsymbol{\tilde{V}_g^k}$, where $\boldsymbol{\tilde{V}_g^k}$ represents the value of the $k$th derivative of $\boldsymbol{\tilde{V}_g}$ reference to $\tau_0$ at $\tau_0=0$. 

The DC Robinson instability is defined by the condition when $\omega_s^2<0$. It can be understood as a condition under which the longitudinal force experienced by a particle becomes defocusing. Eq.~\ref{eq:2.H.1} gives the DC Robinson stability condition explicitly
\begin{equation}
\label{eq:2.H.1}
\sum_n \omega_{g,n}  \boldsymbol{\Im[ \tilde{V}_{g,n}]} >0.
\end{equation}
Meanwhile, $ \sum_n \boldsymbol{\Re[\tilde{V}_{g,n}}]>0$ is also required to compensate for the energy loss due to both synchrotron radiation and the beam-induced voltage~\cite{MIYAHARA1987518}. 

f only one cavity is present, $n=1$,  Eq.~\ref{eq:2.H.1} reduces to
\begin{equation}
\label{eq:2.H.2}
\boldsymbol{\Im[ \tilde{V}_g}] = \boldsymbol{\Im[ \tilde{V}_c} - \boldsymbol{\tilde{V}_b}] =  \boldsymbol{\Im}[ |\boldsymbol{\tilde{V}_c}|e^{i\phi_s} + V_{br} \cos\psi e^{i\psi}]>0, 
\end{equation}
where $\phi_s$ is the synchronous phase. From Eq.~\ref{eq:2.H.2}, the DC Robinson stability criterion~\cite{ng2006physics} can be obtained as 
\begin{equation}
\label{eq:2.H.3}
|\boldsymbol{\tilde{V}_c}|\sin\phi_s + V_{br} \cos\psi \sin\psi > 0.
\end{equation}

Again, when the beam-induced voltage is ignored and denote the generator voltage as $\boldsymbol{\tilde{V}_{g0}}$, following the same derivation as for Eq.~\ref{eq:2.C.6}, the synchrotron frequency $\omega_{s0}$ can similarly be defined as 
\begin{equation}
\label{eq:2.C.8}
\omega_{s0}^2 = 2\frac{e \eta }{T_0E_0 \beta^2} \frac{\omega_g}{2!} \boldsymbol{\Im[ \tilde{V}_{g0}}]. 
\end{equation}

The difference in the particle dynamics between the cases with and without the static beam-induced voltage can be determined by directly comparing $\boldsymbol{\tilde{V}_{g0}}$ and $\boldsymbol{\tilde{V}_{g}}$. The static beam-induced voltage introduces an additional energy shift proportional to $\Re[\boldsymbol{\tilde{V}_{b}}]=\Re[\boldsymbol{\tilde{V}_{g0}} - \boldsymbol{\tilde{V}_{g}}]$ and a shift in the synchrotron oscillation frequency proportional to $\Im[\boldsymbol{\tilde{V}_{b}}]=\Im[\boldsymbol{\tilde{V}_{g0}} - \boldsymbol{\tilde{V}_{g}}]$. 
By explicitly separating the real and imaginary parts, we obtain
\begin{equation}
\label{eq:2.C.9}
\begin{aligned}
%\omega_{s0}^2 - \omega_{s}^2 = 2\frac{e \eta }{T_0E_0 \beta^2} \frac{\omega_g}{2!} \boldsymbol{\Im[ \tilde{V}_{b}}] &= -\frac{e \eta }{T_0E_0 \beta^2} \frac{V_{br} \omega_g \sin2\psi }{2!}, \\
%\boldsymbol{|\tilde{V}_{g0}|} \cos\phi_{g0} - \boldsymbol{|\tilde{V}_{g}|} \cos\phi_{g} = \boldsymbol{\Re[ \tilde{V}_{b}}] &= -V_{br}  \cos^2\psi
&\boldsymbol{|\tilde{V}_{g0}|} \cos\phi_{g0} - \boldsymbol{|\tilde{V}_{g}|} \cos\phi_{g}  = -V_{br}  \cos^2\psi \\
&\omega_{s0}^2 - \omega_{s}^2  = -\frac{e \eta }{T_0E_0 \beta^2} \frac{V_{br} \omega_g \sin2\psi }{2!}, 
\end{aligned}
\end{equation}
where $\phi_{g0}$ and $\phi_{g}$ are the phases of the generator voltages without and with the static beam-induced voltage.

\subsection{Beam-induced voltage at side-bands: the dynamic term}\label{section:2.D} 
The longitudinal beam signal~\cite{byrd1995spectral} at the location of the impedance element is
\begin{equation}
\label{eq:2.D.1}
\boldsymbol{I}_b(t) = - e N \sum_{k=-\infty}^{k=\infty} \sum_{m=0}^{M-1} \boldsymbol{\rho}(t) \delta[t- k T_0 - \frac{m }{M} T_0 + \tau_{m}\cos(\omega_s(k+\frac{m}{M})T_0)],  
\end{equation}
where $m$ is the bunch index varying from $0$ to $M-1$, $\tau_m$ is the oscillation amplitude of the bunch $m$.  Correspondingly, the beam spectrum in the frequency domain can be expressed as   
\begin{equation}
\label{eq:2.D.2}
\begin{aligned}
\boldsymbol{I}_b(\omega) = - e N \omega_0 \sum_{\mu=0}^{M-1} \sum_{l=-\infty}^{l=\infty} (-i)^l J_l(\omega \tau_{\mu}) \sum_{p=-\infty}^{p=\infty}  \boldsymbol{\rho}(\omega) &\delta[\omega-p M \omega_0-\mu  \omega_0 - \Omega_{\mu,l}]  
\end{aligned}
\end{equation}
where $J_l$ is the $l$th-order Bessel function of the first kind, $\tau_{\mu}$ and $\Omega_{\mu,l}$ are the oscillation amplitude and frequency of the CBM $\mu$.  Then, the beam-induced voltage in the frequency domain is 
\begin{equation}
\label{eq:2.D.3}
\begin{aligned}
\boldsymbol{V_b}(\omega)&= - e N \omega_0 \sum_{\mu=0}^{M-1} \sum_{l=-\infty}^{l=\infty} (-i)^l J_l(\omega' \tau_{\mu}) \sum_{p=-\infty}^{p=\infty}  \boldsymbol{\rho}(\omega')  \boldsymbol{Z}(\omega') \end{aligned}
\end{equation}
where $\omega'= p M \omega_0 + \mu  \omega_0 + \Omega_{\mu,l}$. Noticeably, Eq.~\ref{eq:2.D.2} and Eq.~\ref{eq:2.D.3} are general expressions, and they reduce to Eq.~\ref{eq:2.B.2} and Eq.~\ref{eq:2.B.3} respectively if we limit the study to the lowest-order CBM ($\mu=0$) and assume that the beam does not perform any synchrotron motion ($l=0$). In other words, Eq.~\ref{eq:2.D.3} contains both the static term, for which the impedances are sampled at the frequencies $\omega'=Mp\omega_0$, and the ``dynamic" terms, for which the impedances are sampled at the synchrotron side-bands $\omega'= p M \omega_0 + \mu \omega_0 + \Omega_{\mu,l}$. These synchrotron sidebands play a dominant role in driving collective beam instabilities. 

The collective beam motion can be studied in either the time or frequency domain, as described in Chapter 4 of Ref.~\cite{AlexChao}. Using a point-charge model, the frequencies of the collective modes can be expressed by Eq.~(4.128) of Ref.~\cite{AlexChao}. It should be noted that the effect of the static term, given on the right-hand side of Eq.~\ref{eq:2.C.9}, has already been incorporated into $\omega_s$. Assuming that the beam follows a Gaussian longitudinal profile with an rms bunch length $\sigma_z$ and considering only the azimuthal mode $l$, the frequencies of the collective modes can be obtained from Eqs.~(6.102) and (6.204) of Ref.~\cite{AlexChao} as
\begin{equation}
\label{eq:2.D.4}
\begin{aligned}
\Omega_{\mu,l}-l\omega_s &= i \frac{2 \pi M N r_0 \eta c^2}{\gamma T_0^2 \omega_s } l \sum_{p=-\infty}^{p=\infty} \frac{\boldsymbol{Z}(\omega')}{\omega'} g_{l}(\omega')g_{l}(\omega')
%\Omega_{\mu,l}^{2}-(l\omega_s)^2 &=i\frac{2\pi M N r_0 \eta c^2}{\gamma T_0^2 \omega_s} l \sum_{p=-\infty}^{p=\infty} \omega' \boldsymbol{Z}(\omega') \\
%\Omega_{\mu,l}^{2}-(l\omega_s)^2 &\sim \frac{l}{\pi}\frac{\Gamma(l+1/2)}{2^l(l-1)!}\frac{M N r_0 \eta c^3}{\gamma T_0 \sigma_z^3}i(\frac{\boldsymbol{Z}(\omega)}{\omega})_{eff} \\
\end{aligned}
\end{equation}
where 
\begin{equation}
\label{eq:2.D.6}
\begin{aligned}
g_l(\omega) = \frac{1}{\sigma_z \sqrt{2 \pi l!}} (\frac{\omega \sigma_z}{\sqrt{2}c})^{l}e^{-\omega^2 \sigma_z^2/2c^2}.
\end{aligned}
\end{equation}
By solving Eq.~\ref{eq:2.D.4}, the collective-mode frequencies can be determined, providing a basis for identifying and classifying the corresponding instabilities. To extend the analysis to non-Gaussian longitudinal distributions in general, a set of coupled perturbation equations based on Vlasov theory must be solved self-consistently~\cite{lindberg2018theory,PhysRevAccelBeams.21.114404,alves2025theoretical}.

\if(false)
\begin{equation}
\label{eq:2.D.5}
\begin{aligned}
 (\frac{\boldsymbol{Z}(\omega)}{\omega})_{eff} &= \frac{\sum_{p=-\infty}^{p=\infty}\boldsymbol{Z}(\omega')/\omega' h_l(\omega')}{\sum_{p=-\infty}^{p=\infty} h_l(\omega')},
\end{aligned}
\end{equation}
where $h_l(\omega)$ is the bunch spectrum power density. For the beam that follows the Gaussian profile,  
\begin{equation}
\label{eq:2.D.6}
\begin{aligned}
h_l(\omega) = (\frac{\omega \sigma_z}{c})^{2l}e^{-\omega^2 \sigma_z^2/c^2}
\end{aligned}
\end{equation}
\fi

\subsection{Coupled bunch instability or AC Robinson instability}\label{section:2.E} 
If we limit the study to the point-charge model and assume that $\Omega_{\mu,l}$ does not deviate significantly from $l\omega_s$, then, considering the lowest-order $l=1$ mode~\footnote{The $l=-1$ mode has the same growth rate but an opposite mode-frequency shift.}, Eq.~\ref{eq:2.D.4} reduces to Eq.~(4.128) of Ref.~\cite{AlexChao}, as explicitly given in Eq.~\ref{eq:2.E.1}:
\begin{eqnarray}\label{eq:2.E.1}
\Omega_{\mu,1} - \omega_{s} &=&  i \frac{M N r_0 \eta}{2 \gamma T_0^2 \omega_{s}} \sum_{p=-\infty}^{\infty} \omega' \boldsymbol{Z}(\omega').
\end{eqnarray} 
This equation shows that there are $M$ CBMs in total. For the $\mu$th CBM, the impedance is sampled at $\omega'= p M \omega_0 + \mu \omega_0 + \omega_s$. Eq.~\ref{eq:2.E.1} is commonly used to obtain a first estimate of the growth rate.

Similarly, $l\neq1$ modes can also be excited, typically leading to bunch-length oscillations and an increase in the energy spread. In a recent study at MAX IV~\cite{PhysRevAccelBeams.27.044403}, the $l=1,2$ modes with CBM indices $\mu=0,\pm 1$ were experimentally measured.

In the cavity design, it is desirable to properly control all $M$ CBMs. Assume that the cavity impedance of the fundamental mode at the generator frequency $\omega_g=h\omega_0$ is given by Eq.~\ref{eq:2.1}. If there is only one bunch in the ring so that $M=1$ and $\mu=0$, then the beam  samples the impedance at the side-bands of every revolution frequency $\omega'= \sum_{p=-\infty}^{\infty} (p \omega_0 +\omega_s)$. If the quality factor of the cavity mode is high, the dominant contributions to Eq.~\ref{eq:2.E.1} come from $\omega'=\pm h \omega_0 + \omega_s$. If the cavity-mode detuning frequency is also larger than the synchrotron frequency $f_s=\omega_s/2\pi$, then the growth rate can be estimated as
\begin{eqnarray}\label{eq:2.E.2}
\begin{aligned}
%1/\tau_{\mu} &\propto  (- h \omega_0 + \omega_s) \boldsymbol{\Re Z}(- h \omega_0 + \omega_s) + (h \omega_0 + \omega_s) \boldsymbol{\Re Z}(h \omega_0 + \omega_s) \\
%&\approx  h \omega_0 (  \boldsymbol{\Re Z}(h \omega_0 + \omega_s) -  \boldsymbol{\Re Z}(h \omega_0 - \omega_s)) \\ 
%&\approx 2\omega_s  h \omega_0   \boldsymbol{\Re Z'}(h \omega_0), \\
1/\tau_{\mu} &\propto 2\omega_s  h \omega_0   \boldsymbol{\Re Z'}(h \omega_0), 
\end{aligned}
\end{eqnarray} 
where $\boldsymbol{'}$ denotes differentiation with respect to frequency. Beam stability requires $\boldsymbol{\Re Z'}(h \omega_0)<0$, which implies $\omega_r<h\omega_0=\omega_g$. However, when $M>1$, $\omega_r<\omega_g$ ensures only the stability of the $\mu=0$ CBM. Other CBMs may still have positive growth rates.

\subsection{Longitudinal mode coupling instability}\label{section:2.F} 
The longitudinal mode coupling instability (LMCI)~\cite{AlexChao,wang1980condition,oide1990longitudinal,karpov2023longitudinal,wang1989robinson,krinsky1983longitudinal} can be excited among different azimuthal modes $l$. In the low-beam-current regim, for a given CBM index $\mu$, the oscillation frequency of a CBM is located at $\omega' = p M \omega_0 + \mu \omega_0 + l\omega_s$ on the beam spectrum. As the beam current increases, the frequency of the collective mode $\Omega_{\mu,l}$ shifts. Instabilities could be excited when any two modes (different $l$ index) merge. A systematic study of the LMCI requires a model based on the Vlasov equation \cite{AlexChao,ng2006physics,alves2025theoretical}.  For a rough estimation, the frequency shift of the $l$ mode can be estimated from the real part of Eq.~\ref{eq:2.D.4} as
\begin{eqnarray}\label{eq:2.F.1}
\begin{aligned}
\Re[\Delta\Omega_{\mu,l}]=\Re[(\Omega_{\mu,l}-l\omega_s)]= 
-\frac{2 \pi M N r_0 \eta c^2}{\gamma T_0^2 \omega_s } l \sum_{p=-\infty}^{p=\infty} \frac{\Im[\boldsymbol{Z}(\omega')]}{\omega'} g_{l}(\omega')g_{l}(\omega')
\end{aligned}
\end{eqnarray}
The beam-current threshold can then be approximately estimated from the condition
\begin{eqnarray}\label{eq:2.F.2}
\begin{aligned}
%\Delta\Omega_1 &= i \frac{M N r_0 \eta}{2 \gamma T_0^2 \omega_{s}} \sum_{\mu=0}^{M}\sum_{p=-\infty}^{\infty} \omega' \boldsymbol{Z}(\omega') \quad  & \omega'= pM\omega_0 + \mu \omega_0 + \omega_s \\
\Re[\Delta\Omega_{\mu,l}] - \Re[\Delta\Omega'_{\mu,l}] \approx \pm \omega_s.
\end{aligned}
\end{eqnarray} 
where $\Re[\Delta\Omega'_{\mu,l}]$ represents the contribution from potential-well distortion (the static beam-induced voltage term) that impedance is sampled at $p M \omega_0 + \mu \omega_0$. Noticeably, the mode frequency shift from Eq.~\ref{eq:2.F.1} is $\propto MN$. If the ring is limited to a low single-bunch charge population $N$, one can increase the number of bunches $M$ to observe the LMCI. 

\subsection{Beam-Robinson mode and Cavity-Robinson mode}\label{section:2.G} 
Returning to Eq.~\ref{eq:2.D.4}, the mode defined there is referred to as the beam-Robinson mode in Refs.~\cite{PhysRevSTAB.4.074401,yamaguchi2023systematic}. However, if we remove the approximation $\Omega_{\mu} \approx \omega_s$, limit the discussion to the point-charge model, and consider only the $l=1$ mode, then the mode frequency follows the equation
\begin{eqnarray}\label{eq:2.G.1}
\Omega^2_{\mu} - \omega_{s}^2 =  i \frac{M N r_0 \eta}{\gamma T_0^2 } \sum_{p=-\infty}^{\infty} \omega' \boldsymbol{Z}(\omega').
\end{eqnarray} 
where $\omega'= p M \omega_0 + \mu \omega_0 + \Omega_{\mu}$. The second solution branch of Eq.~\ref{eq:2.G.1} is referred to as the cavity-Robinson mode~\cite{yamaguchi2023systematic} or the D-mode~\cite{PhysRevAccelBeams.27.064402}. If the impedance is sharply peaked near the resonant frequency $\omega_g=h\omega_0$, then the dominant contribution comes from the $\mu$th CBM that satisfies $p M + \mu = h$. Taking this particular mode $\mu$ as an example, Eq.~\ref{eq:2.G.1} can be approximated as
\begin{eqnarray}\label{eq:2.G.2}
\Omega^2_{\mu} - \omega_{s}^2 =  i \frac{M N r_0 \eta}{\gamma T_0^2}[(h\omega_0+\Omega_{\mu}) \boldsymbol{Z}(h\omega_0+\Omega_{\mu}) - (h\omega_0 - \Omega_{\mu}) \boldsymbol{Z}(h\omega_0-\Omega_{\mu})].
\end{eqnarray}
Eq.~\ref{eq:2.G.2} can be further simplified by applying the approximation given in Eq.~\ref{eq:4.A.3}. From there, a characteristic equation of $\Omega_{\mu}$ can be established and solved analytically. Further details can be found in Refs.~\cite{yamaguchi2023systematic,PhysRevAccelBeams.27.064402}.  

\subsection{Numerical method in particle tracking }\label{section:2.I}
In particle tracking, the key is to get the right beam-induced voltage $\boldsymbol{V_{b}}(t)$ to kick the particles.  Starting from the fundamental theory of beam loading, the procedure can be significantly simplified. Assume that there is a bunch with charge $q$ passing through the cavity at time $t-\Delta t$, then the beam-induced voltage follows 
\begin{equation}
\label{eq:2.I.1}
\boldsymbol{V_{b}}(t) = (\boldsymbol{V_{b}}(t - \Delta t) + \boldsymbol{V_{b0}}/2)\exp(\alpha \Delta t), \quad
\alpha = -\frac{1}{\tau_{f}}(1-i \tan{\psi}).
\end{equation}
The phase of the term $\boldsymbol{V_{b0}}$ is $\pi$, and its amplitude is $q \omega_{r} R_{L} / Q_{L}$. Eq.~\ref{eq:2.I.1} shows that the accumulated beam-induced voltage $\boldsymbol{V_{b}}(t)$ receives an impulse $\boldsymbol{V_{b0}}/2$ whenever a charged bunch passes through the cavity, and then decays and rotates by a factor of $\exp(\alpha \Delta t)$ until the next bunch arrives. For bunches with a finite length, each bunch can be divided into time bins, and each bin can be treated as a zero-length microbunch. Combined with the generator dynamics described by Eq.~\ref{eq:2.A.2}, the electron particles can then be tracked self-consistently.

\section{Estimation of instabilities in Petra-IV}\label{section:3}
\begin{figure}
\centering
\includegraphics[width=1\textwidth]{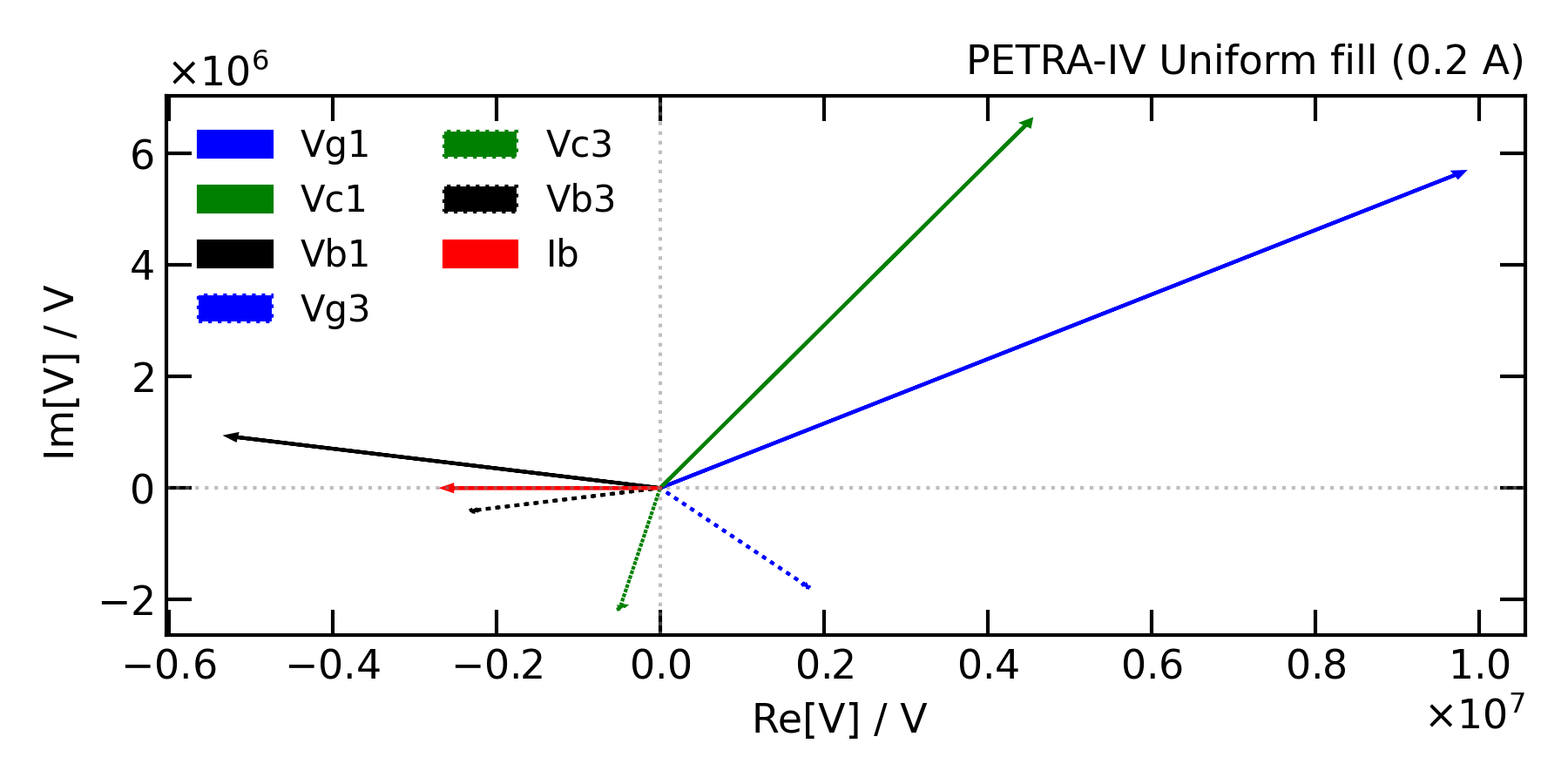}
\caption{\label{fig:3.1} The static beam-induced voltage $\boldsymbol{\tilde{V}_b}$, cavity voltage $\boldsymbol{\tilde{V}_c}$ and generator voltage $\boldsymbol{\tilde{V}_g}$ in the $e^{i\omega_g}$ frame. The total beam current is assumed as 200 mA.}
\end{figure}

In this section, we estimate the instability growth rate of the dual active normal-conducting RF system under the beam conditions of the timing operation mode of PETRA-IV. The parameters of the fundamental modes of the main and harmonic cavities are listed in Tab.~\ref{tab:2}. The ideal bunch-lengthening condition requires $\sum_n\Re[\boldsymbol{\tilde{V}{c,n}}]=U_0/e$ to compensate for the synchrotron-radiation energy loss $U_0$, and $\sum_n a{1,n} =\sum_n a_{2,n}=0$ to produce a flat RF potential well. The voltage of the main cavity is set to 8 MV. with the help of the ideal-bunch lengthening condition, the required harmonic cavity voltage can be determined accordingly.  Fig.~\ref{fig:3.1} shows the phasors of static beam-induced voltage $\boldsymbol{\tilde{V}_b}$, cavity voltage $\boldsymbol{\tilde{V}_c}$ and generator voltage $\boldsymbol{\tilde{V}_g}$ in the frame rotating at the main-cavity RF frequency ($e^{i\omega_g}$) for the brightness operation mode. Tab.~\ref{tab:3} shows their values in the timing and brightness operation schemes explicitly.

\begin{table}[htp]   
\caption{Nominal values of the voltage phasor $\boldsymbol{\tilde{V}_b}$, $\boldsymbol{\tilde{V}_c}$ and $\boldsymbol{\tilde{V}_g}$.}
\centering
\begin{tabular}{|l|r|r|r|r|r|r|}\hline \hline
cavity harm & $Abs(\boldsymbol{\tilde{V}_{g}})$ & $Abs(\boldsymbol{\tilde{V}_{b}})$ 
 & $Abs(\boldsymbol{\tilde{V}_c})$ & $Arg(\boldsymbol{\tilde{V}_g})$ 
 & $Arg(\boldsymbol{\tilde{V}_b})$ & $Arg(\boldsymbol{\tilde{V}_c})$ \\
\hline
\multicolumn{7}{|c|}{\textbf{Timing mode 80b/80mA}}  \\ \hline
 n=1  & 9.15$\times 10^6$  & 2.14$\times 10^6$ & 8$\times 10^6$  & 0.727 & 2.968 & 0.939  \\ 
\hline
 n=3  & 2.03$\times 10^6$  & 9.45$\times 10^5$ & 2.22$\times 10^6$ & -1.37 & -2.967 & -1.81  \\ \hline
 \multicolumn{7}{|c|}{\textbf{Brightness mode 1920b/200mA}}  \\ \hline
 n=1  & 11.43$\times 10^6$  & 5.35$\times 10^6$ & 8$\times 10^6$  & 0.504 & 2.968 & 0.939  \\ 
\hline
 n=3  & 2.51$\times 10^6$  & 2.36$\times 10^6$ & 2.22$\times 10^6$ & -0.768 & -2.967 & -1.81 \\
\hline
\end{tabular}
\label{tab:3}
\end{table} 

\begin{figure}
\centering
\includegraphics[width=1\textwidth]{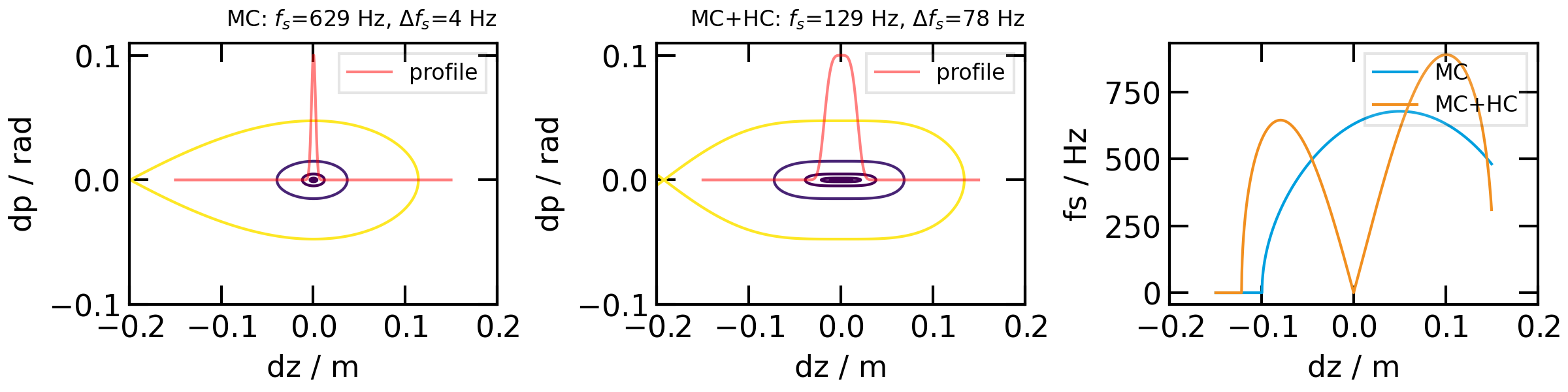}
\caption{\label{fig:3.2} Comparison of beam profile, the longitudinal phase space tori and longitudinal oscillation $f_s$ as a function of beam offset $dz$ in the double RF and single RF system.}
\end{figure}

Assuming that the beam is in a steady state, Fig.~\ref{fig:3.2} shows a comparison of the beam characteristics between the single- and dual-RF systems, where Eqs.~\ref{eq:2.C.11} and \ref{eq:2.C.6} are used. In the dual-RF system, the flat-potential-well condition leads to bunch lengthening by a factor of 5. The Hamiltonian tori in the longitudinal phase space are significantly distorted, and the corresponding oscillation frequency along each trajectory is also significantly reduced within the beam. The more pronounced variation in $f_s$ also leads to a broader oscillation-frequency spread. By performing a weighted average over the beam profile, it is found that the average longitudinal oscillation frequency $\bar{f_s}$ decreases from 630~Hz to 130~Hz, while its spread $\Delta{f_s}$ increases from 4~Hz to 80~Hz. Due to the possible overlap between the incoherent synchrotron-frequency spread and the frequencies of the longitudinal collective modes, beam stability can be enhanced by Landau damping~\cite{herr2016introduction,ng2006physics}. A detailed analysis of the corresponding Landau stability diagram is beyond the scope of this manuscript.

\subsection{DC Robinson instability}\label{section:3.0}
With the help of Eq.~\ref{eq:2.H.1}, the RF parameters in Tab.~\ref{tab:2} can be checked and they ensure the DC Robinson stable condition.

\subsection{Coupled bunch instability}\label{section:3.1}
\begin{figure}
\centering
\includegraphics[width=1\textwidth]{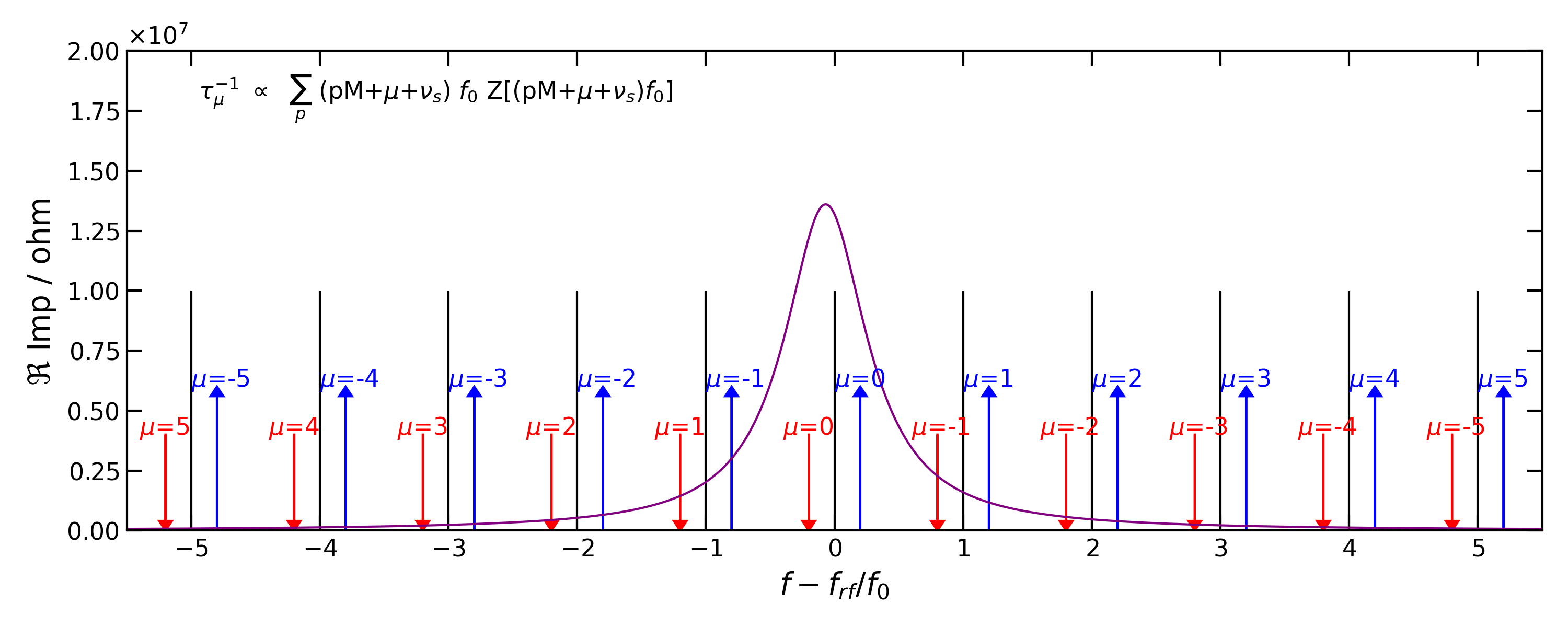}
\caption{\label{fig:3.3} Qualitative illustration of beam spectrum and CBM numbering for a uniform beam fill pattern in PETRA-IV along the real part of the impedance of the fundamental mode from the main cavity. For each sideband, blue or red, the direction of the arrows indicates excitation or damping of the CBM. The negative CBM index $\mu$ is equivalent to mode $h+\mu$.}
\end{figure}

\begin{figure}
\centering
\includegraphics[width=1\textwidth]{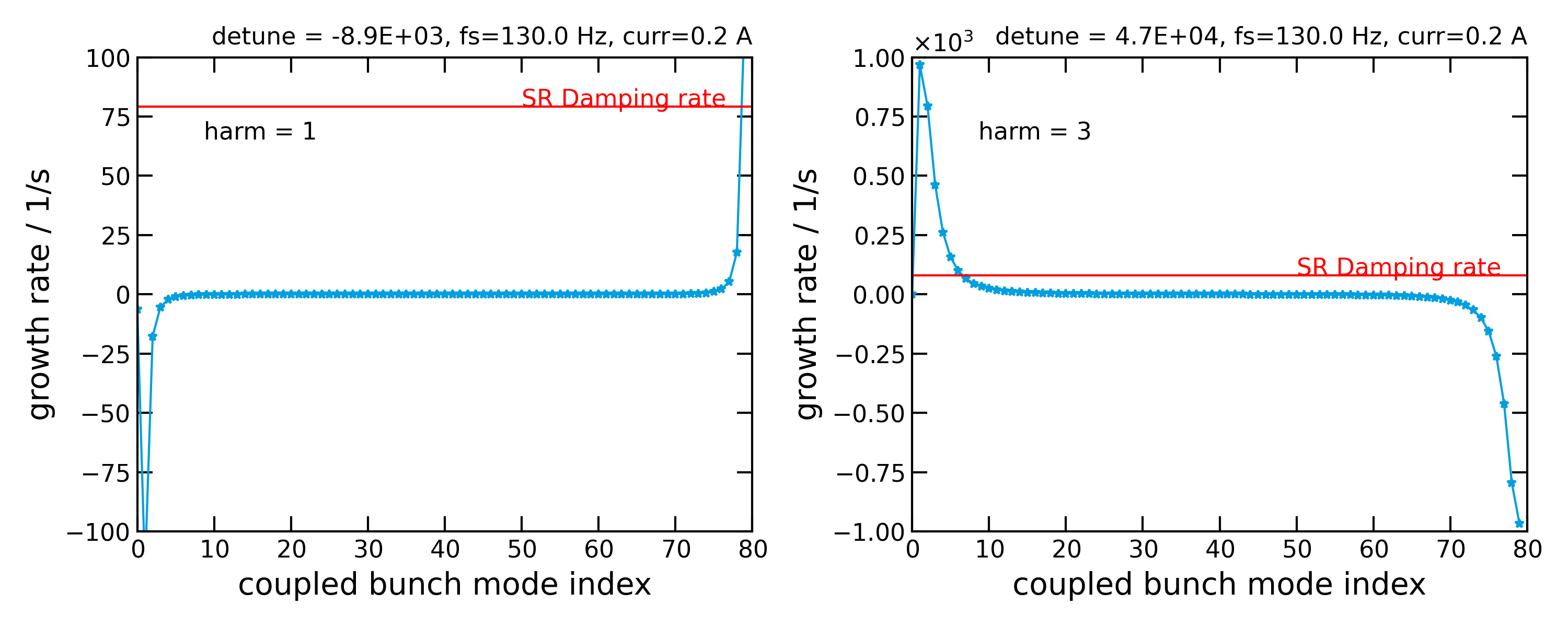}
\caption{\label{fig:3.4}The CBM growth rate from the fundamental mode of the main (left) and harmonic (right) cavities when the ring is assumed to be uniformly filled by 80 bunches with a total beam current 200 mA, the longitudinal frequency is set as $f_s=130$ Hz.}
\end{figure}

Using Eq.~\ref{eq:2.D.4}, we can estimate the longitudinal coupled-bunch instabilities for various azimuthal modes $l$. Fig.~\ref{fig:3.3} qualitatively illustrates the beam spectrum and CBM numbering for a uniform beam-filling pattern, along with the real part of the impedance. The growth rates of the CBM for the azimuthal mode index $l=1$ are shown in Fig.~\ref{fig:3.4}, where the ring is uniformly filled with 80 bunches at a total beam current of 200~mA. With negative and positive detuning frequencies for the main and harmonic cavities, respectively, the $\mu=-1$ and $\mu=1$ CBM instabilities are excited, respectively. The growth rates of the $l\neq1$ modes are much lower and are therefore not shown. Notably, the growth rate driven by the harmonic cavity is much higher than the natural damping rate, indicating that an additional damping mechanism is required to stabilize the beam, such as LLRF.

\if(false)
\subsection{Longitudinal mode coupling instability}\label{section:3.2}
With help from Eq.~\ref{eq:2.F.2}, the beam current threshold due to longitudinal mode coupling can be estimated. The frequency shift of the $l$ mode due to the both cavities are much smaller than $\omega_s/2$. We do not expect the longitudinal modeling coupling instability can be excited.   
\fi

\section{Low-level RF feedback}\label{section:4}
Low-level radio-frequency feedback is indispensable for maintaining precise amplitude and phase control of the accelerating cavities and preserving the required longitudinal bucket area, and it typically responds on a timescale shorter than the synchrotron period $1/f_s$. Properly designed LLRF feedback can compensate for imperfections such as detuning caused by microphonics, Lorentz-force effects, and beam-induced voltage, among others, thereby ensuring stable machine operation~\cite{fox2010lessons,drago2003longitudinal,kallakuri2022coupled}. LLRF feedback can also mitigate longitudinal coupled-bunch instabilities by significantly reducing the effective closed-loop impedance experienced by the beam. According to the signal delay in the feedback loop, LLRF feedback can be classified into short-delay feedback, also referred to as direct RF feedback (DRF-FB), and long-delay feedback. In general, DRF-FB reduces the impedance over a broad frequency range, whereas long-delay feedback reduces the impedance around the revolution harmonics. In PETRA-IV, both DRF-FB and long-delay feedback are included in the baseline design for both the main and harmonic cavities. The DRF-FB employs a proportional-integral-derivative (PID) controller, and the delay time of the long-delay feedback is set to one turn. In this section, we introduce the LLRF feedback system in general and use a description in the Laplace $s$-domain for convenience. The corresponding frequency-domain representation can be obtained by setting $s=i\omega$.

\subsection{Low-level RF feedback}\label{section:4.A}
\begin{figure}[!ht]
\centering
\includegraphics[width=0.8\textwidth]{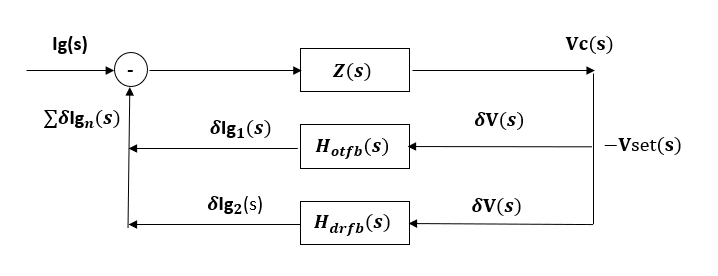}
\caption{\label{fig:4.1} Beam cavity RF-feed back control loops.}
\end{figure}

Figure~\ref{fig:beam_cavity_feedback} shows the coupled dynamics among the generator, the beam, and the cavity fundamental mode. The fundamental mode of the cavity responds to both the generator current $\mathbf{I_g}$ and the beam current $\mathbf{I_b}$. The LLRF feedback is designed to stabilize the cavity voltage experienced by the beam. The LLRF feedback control loop can be simplified as shown in Fig.~\ref{fig:4.1}, and the closed-loop transfer function of the fundamental mode can then be expressed as
\begin{equation}\label{eq:4.A.1}
Z_{cl}(s) = \frac{Z(s)}{ 1+ H(s) Z(s)} = \frac{Z(s)}{ 1+ D(s) G(s) Z(s) e^{i \phi}}, 
\end{equation}
where $Z(s)$ is the cavity transfer function when the control loop is open, $H(s)$ is the overall feedback transfer function, which can be expressed as $H(s)=D(s)G(s)e^{i\phi}$, with $G(s)$ representing the transfer function of the controller, $D(s)=\exp(-s \tau)$ representing the delay function with a delay time $\tau$, and $\phi$ representing the phase shift. A larger open-loop gain $|H(s)Z(s)|$ leads to a greater reduction in the closed-loop impedance.

In the following, we denote the angular bandwidth of the RF mode by $\sigma=\omega_{1/2}$. The cavity-impedance transfer function given by Eq.~\ref{eq:2.1} can be expressed in the Laplace $s$-domain as
\begin{equation}\label{eq:4.A.2}
%Z(s) = \frac{2 R_{L} \sigma s}{s^2 + 2\sigma s + \omega_r^2}
 Z(s) = \frac{-2 R_{L} \sigma s}{s^2 - 2\sigma s + \omega_r^2}.
\end{equation}
The transfer function in the vicinity of the driving frequency $\omega_{rf}$ is 
\begin{equation}\label{eq:4.A.3}
Z(s\pm i \omega_{rf}) =  \frac{R_{L} \sigma }{-s + \sigma \pm  i \sigma \tan\psi},
\end{equation}
where $\psi$ is the detuning phase at $\omega=\omega_{rf}$. Since $\omega_{\rm rf}$ is the natural choice for the carrier frequency of the feedback system, unless otherwise specified, all frequencies in the following are defined relative to $\omega_{\rm rf}$, and the same convention is used for the impedance and transfer functions.

\subsection{Delay $D(s)$}\label{section:4.B}
To assess the impact of the delay $D(s)$~\cite{boussard1985control} on the DRF-FB, we assume, for simplicity, that two conditions are satisfied: 1) the controller $G(s)$ behaves smoothly and has no singularities around the resonant frequency $\omega_r$; 2) the cavity detuning $\psi$ and the phase shift $\phi$ are both zero. Then, the open-loop cavity impedance around the resonant frequency can be expressed as, 
\begin{equation}\label{eq:4.B.1}
Z(\omega) = \frac{R_{L} }{1 - i \frac{\omega}{\sigma}} \approx
\begin{cases}
R_{L}, & \text{$\omega \ll \sigma$ }.\\
\frac{R_{L} }{ - i \frac{\omega}{\sigma}}, & \text{$\omega \gg \sigma$ }
\end{cases}.
\end{equation}
The corresponding closed-loop impedance is
\begin{equation}\label{eq:4.B.2}
Z_{cl}(\omega) = \frac{Z(\omega)}{ 1+ \exp(-i\omega \tau) G(\omega) Z(\omega)} \approx
\begin{cases}
\frac{R_{L}}{1+GR_{L}}, & \text{$\omega \ll \sigma$ }.\\
\frac{R_{L}}{ -i \frac{\omega}{\sigma} +  \exp(-i\omega \tau) G(\omega) R_{L} }, & \text{$\omega \gg \sigma$ }
\end{cases}.
\end{equation}
Within the cavity bandwidth $\sigma$, a larger open-loop gain $|GR_{L}|$ at resonant frequency leads to a significant impedance reduction.  Outside the bandwidth $\sigma$, the cavity is purely reactive with a phase shift $-\pi/2$. A classic indicator of stability, for a feedback loop, is the phase margin defined as the amount by which the phase of the open-loop response exceeds $\pi$ when the modulus of its gain is one~\cite{baudrenghien2001low}. If the delay introduces an additional  $\pi/4$ phase shift so that $\omega \tau=\pi/4$, to keep a phase safety margin of $\pi/4$, the magnitude of the open-loop gain must not exceed unity at this frequency, such that
\begin{equation}\label{eq:4.B.3}
\begin{aligned}
&|G(\pi/4\tau)Z(\pi/4\tau)| \le 1 \\
&\Rightarrow |G(\pi/4\tau) \frac{R_{L} }{1 - i \frac{\pi}{4\tau\sigma}}| \approx G \frac{4R_{L} \sigma \tau }{\pi}  \le 1 \\
&\Rightarrow G R_{L} \le \frac{\pi}{4 \sigma \tau} = G_{max}  R_{L} 
\end{aligned}
\end{equation}
This condition gives an upper limit for $|G(\omega)Z(\omega)|$ and a minimum value of impedance experienced by the beam at the frequency $\omega=0$
\begin{equation}\label{eq:4.B.4}
\begin{aligned}
R_{min}  =  \frac{R_{L}}{1+G_{max}R_{L}}  \Rightarrow
\frac{R_{min}}{R_{L}} =\frac{1}{1+G_{max}R_{L}}  \approx \frac{4 \sigma \tau}{\pi}  = \Gamma \tau.
\end{aligned}
\end{equation}

With cavity parameters listed in Tab.~\ref{tab:1}, to reduce the closed-loop impedance by a factor of 4, the maximum allowable delay times $\tau$ for the main and harmonic cavities are approximately 600~ns and 120~ns, respectively.

\subsection{PID controller}\label{section:4.C}
The PID controller is commonly used in control systems, with a transfer function
\begin{equation}\label{eq:4.C.1}
G(s)= (K_p + K_i/s + K_d s)  \\
\end{equation}
Here, $G(s)$ is normalized by $R_L$ and is therefore dimensionless. Substituting Eq.~\ref{eq:4.C.1} into Eq.~\ref{eq:4.A.1}, the closed-loop transfer function can be obtained accordingly. In practice, the derivative term $K_d$ is often set to zero. Near the resonant frequency $f=f_{rf}$, the closed-loop impedance is reduced to $R_{L}/(1+K_p)$. The $K_p$ term, together with the time delay $D(s)$, introduces additional resonant structures in both the amplitude and phase responses. Overshoot around $f_{rf}$ in the amplitude response normally indicates a potential source of beam instability. The $K_i$ term produces a pronounced notch at $s=0$. The larger $K_i$ is, the wider the notch bandwidth becomes. The pole at $s=0$ introduced by the integral term $K_i$ violates one of the assumptions made in Sec.~\ref{section:4.B}, and therefore care must be taken when using Eq.~\ref{eq:4.B.4} for an estimate. 

\subsection{DRF-FB in {$z$} domain}\label{section:4.D}
In the time domain, a transfer function $H(s)$ can generally be represented by an infinite-impulse-response (IIR) filter: 
\begin{equation}\label{eq:4.D.1}
\begin{aligned}
a_0 y(n) &=& -\sum_{i=1}^{Q}a_i y(n-i) + \sum_{j=0}^{P}b_j x(n-j).
\end{aligned}
\end{equation}
The coefficients $a_i$ and $b_j$ are the filter coefficients, and $Q$ and $P$ are corresponding filter orders. Normally, $a_0$ is set as 1. Then, the transfer function can be obtained by performing the $z$ transform~\cite{Z-Transform} 
\begin{equation}\label{eq:4.D.2}
H(z) = \frac{Y(z)}{X(z)} = \frac{Z[y(n)]}{Z[x(n)]} =\frac{\sum_{j=0}^{P}b_j z^{-j}}{1 + \sum_{i=1}^{Q}a_i z^{-i}}.
\end{equation}
$H(z)$ plays the same role in the discrete-time domain as $H(s)$ does in the continuous-time domain in Eq.~\ref{eq:4.A.1}. If we denote the feedback sampling and processing time by \(T_s\), the continuous-time and discrete-time variables are related by \(z=e^{sT_s}\). For numerical implementation, we employ the bilinear transformation, for which the mapping between the $s$ and $z$ domains is
\begin{equation}\label{eq:4.D.3}
s = \frac{2}{T_s}\frac{1-z^{-1}}{1+z^{-1}},
\end{equation}
which can then be applied in the numerical simulation. If we note the delay $\tau=m T_s$, the transfer function of the DRF-FB~\cite{wang2023rf} can be obtained as
\begin{equation}
\label{eq:4.D.4}
H_{DRFB}(z)=D(z)G(z)e^{i\phi} =\frac{(K_p + K_i T_s /2) z^{-m}   + (-K_p + K_i T_s /2) z^{-(m+1)}}{(1-z^{-1})}e^{i\phi}.
\end{equation}
Comparing Eq.~\ref{eq:4.D.4} with Eq.~\ref{eq:4.D.1}, we obtain
\begin{equation}\label{eq:4.D.5}
\begin{aligned}
a_0&=1, \quad a_1=-1; \\
b_0&=...=b_{m-1}=0, \quad b_m =(K_p + K_i T_s /2) e^{i\phi}, \quad  b_{m+1} = (-K_p + K_i T_s /2) e^{i\phi}
\end{aligned}
\end{equation}
%Then real and imaginary part of the $b_m$ composes of the In-Phase and Quadrature feedback digital filters.

\subsection{Long-delay feedback: One-turn delay feedback}\label{section:4.E}
Besides the DRF-FB, the one-turn-delay feedback (OTD-FB) is also widely applied to suppress the effective impedance beam samples~\cite{baudrenghien2012lhc}. In the $s$ domain, the transfer function of the OTD-FB is 
\begin{equation}\label{eq:4.E.1}
H_{OTDFB}(s) = G_0 \frac{(1-\alpha)e^{-s T_0 }}{1-\alpha e^{- s T_0 }} =  G_0 \frac{ (1-\alpha) e^{- s m_s T_s }}{1-\alpha e^{-s m T_s }},
\end{equation}
where $G_0$ is normalized by $R_L$ to give a dimensionless OTD-FB gain, $T_0=1/f_0$ is the one-turn revolution period, $\alpha$ is the comb-filter coefficient, and $m_s=T_{0}/T_s$. The transfer function in the $z$-domain can then be expressed as
\begin{equation}\label{eq:4.E.2}
H_{OTDFB}(z) = G_0  \frac{(1-\alpha)z^{-m_s}}{1-\alpha z^{-m_s}},
\end{equation}
which gives
\begin{equation}\label{eq:4.E.3}
\begin{aligned}
a_0&=1,  \quad a_{m_s}=-\alpha; \\
b_0&=...=b_{m-1}=0, \quad b_{m_s} =G_0 (1-\alpha)
\end{aligned}
\end{equation}

From Eq.~\ref{eq:4.E.2}, it can be seen that the gain at the passband, $z^{-m_s}=1$, is $G_0$. The OTD-FB creates notches in $Z_{cl}(s)$ at every revolution harmonic, and the $-3$dB bandwidth can be approximated as
\begin{equation}\label{eq:4.E.4}
\Delta f_{-3db} \approx \frac{1}{2\pi} (1-\alpha) f_0. 
\end{equation}
A smaller value of $1-\alpha$ results in a narrower bandwidth. As a result, within the OTD-FB filter bandwidth, the effective impedance at every revolution harmonic is reduced by a factor of $1+G_0$. Thus, if the bandwidth of the OTD-FB covers the synchrotron sidebands, both the static beam-induced voltage and the coupled-bunch instabilities can be suppressed, which requires $\Delta f_{-3db}>f_s$. At the condition $z^{-m_s}=-1$, magnitude of the gain reaches its minimum value
\begin{equation}\label{eq:4.E.5}
H^{min}_{OTDFB} = G_0 \frac{1-\alpha}{1+\alpha} \approx = G_0 \frac{1-\alpha}{2}, 
\end{equation}
and the phase response is flipped by 2$\pi$. To maintain a reasonable stability margin, the gain of the OTD-FB should be significantly below 1 at $z^{-m_s}=-1$ due to the Nyquist stability criteria. For a gain margin of approximately 10~dB, the following condition should be satisfied~\cite{baudrenghien2001low}.  
\begin{equation}\label{eq:4.E.6}
H^{min}_{OTDFB} = G_0 \frac{1-\alpha}{2} < 1/3.
\end{equation}

For further improvement, a double-peaked comb filter can be used, with the transfer function given by
\begin{equation}\label{eq:4.E.7}
H_{dcomb}(s) = G_{dcomb} \frac{(1-e^{-sT_0})e^{-s(T_0-\tau_g)}}{(1-\alpha^{-sT_0-i \omega_sT_0})(1-\alpha e^{-sT_0 + i \omega_sT_0})},
\end{equation}
and $\tau_g$ is an additional delay~\cite{baudrenghien2001low,Pedersen:244817}. This implementation requires signals from the previous two turns to construct the feedback signal.

\if(false)
Note the longitudinal phase advance as  $2\pi \nu_s$, and set $\tau_g=0$, correspondingly, the transfer in $z$ domain can be simplified to 
\begin{equation}\label{eq:4.E.8}
H_{dcomb}(z) = G_{dcomb} \frac{z^{-m_s}-z^{-2m_s}}{(1-\alpha z^{-m_s}e^{-i2\pi \nu_s}) (1-\alpha z^{-m_s} e^{i 2\pi \nu_s})}.
\end{equation}
\fi

\subsection{Implementation RF feedback in particle tracking}\label{section:4.F}
Here, we give a short discussion on how to simulate the LLRF feedback consistently together with particle tracking.  Back  to Eq.~\ref{eq:4.D.1}, the generator current error signal $\mathbf{\delta \tilde{I}_g}(n)$ plays as $y(n)$ and the cavity voltage error signal $\mathbf{\delta \tilde{V}_c}(n)$ plays as $x(n)$.  With the previous sampled generator current $\mathbf{\tilde{I}_{g}}(n-i)$ and voltage $\mathbf{\delta \tilde{V}_c}(n-j)$, the generator current can be updated at every feedback sampling $T_s$. Thereafter, the generator dynamics can be advanced by solving Eq.~\ref{eq:2.A.2},  Fig.~\ref{fig:beam_cavity_feedback}. By default, the coefficient $b$ in Eq.~\ref{eq:4.D.1} is normalized by the cavity-loaded impedance $R_L$ to ensure the consistency of the dimensions.

\section{Application in PETRA-IV}\label{section:5}
In this section, we study how longitudinal beam motion can be stabilized by LLRF feedback in PETRA-IV. The brightness operation mode is adopted as the default beam condition. As shown in Fig.~\ref{fig:3.4}, the harmonic cavity is the dominant source of instability. In the following, we first introduce the LLRF feedback specifications. We then present an analysis of three feedback configurations. Finally, we verify through numerical simulations that the beam can be stabilized by the LLRF feedback loops. The simulations are performed using Elegant~\cite{elegantWebPage}, with each bunch represented by $10^4$ macroparticles as a compromise between accuracy and computational efficiency.

\subsection{LLRF feedback specifications}\label{section:5.A}
As discussed, the DRF-FB reduces the effective impedance experienced by the beam over a broad frequency range, whereas the OTD-FB reduces the effective impedance around each revolution harmonic. In PETRA-IV, the nominal operating frequency of the RF feedback system is $f_{rf,1}/4 \approx 125$~MHz, corresponding to a sampling time $T_s\approx8$ ns. The minimum time delay $\tau$ of the DRF-FB is estimated to be 1000 ns~\footnote{Limited by the hardware configuration.}. Based on these specifications, a preliminary controller is designed for both the main and harmonic cavities as
\begin{equation}
\label{eq:5.1}
\tau = 125 T_{s}, \quad \phi=0, \quad  K_p = K_p, \quad K_i=0. 
\end{equation}
Since the OTD-FB creates notches at every harmonic frequency, we set $K_i$ to zero and leave $K_p$ as a free parameter for optimization. 

As shown in Sec.~\ref{section:3}, when the cavities operate under the designed conditions, the longitudinal oscillation frequency is approximately $f_s=130/630$~Hz when the harmonic cavity is turned on/off, respectively. Taking these frequencies into account, we choose $\alpha=31/32\approx0.968$ for the OTD-FB, which gives a bandwidth of $\Delta f_{-3db}=647$~Hz. According to Eq.~\ref{eq:4.E.6}, the maximum allowable OTD-FB gain is 21.

\subsection{Closed-loop impedance and CBM growth rate}\label{section:5.B}
\subsubsection{DRF-FB: PID controller}\label{section:5.B.1}
\begin{figure}
\includegraphics[width=1\textwidth]{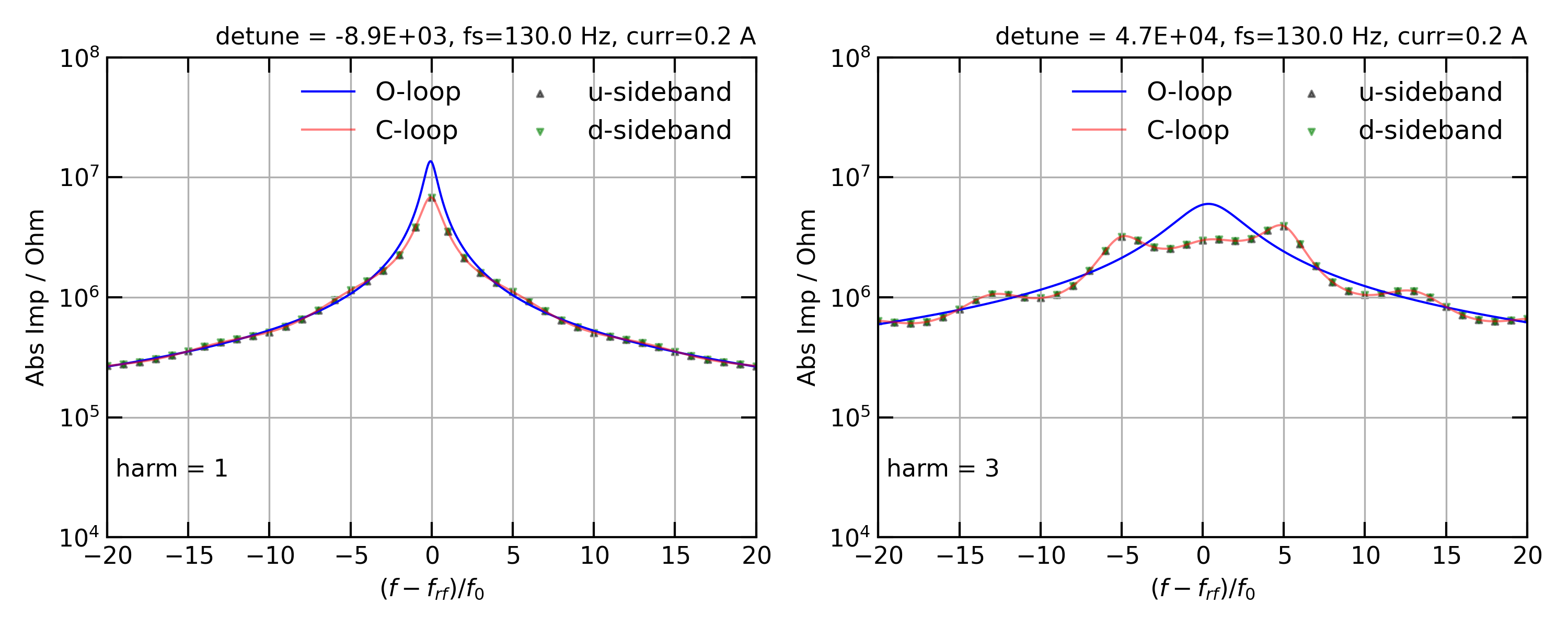}
\includegraphics[width=1\textwidth]{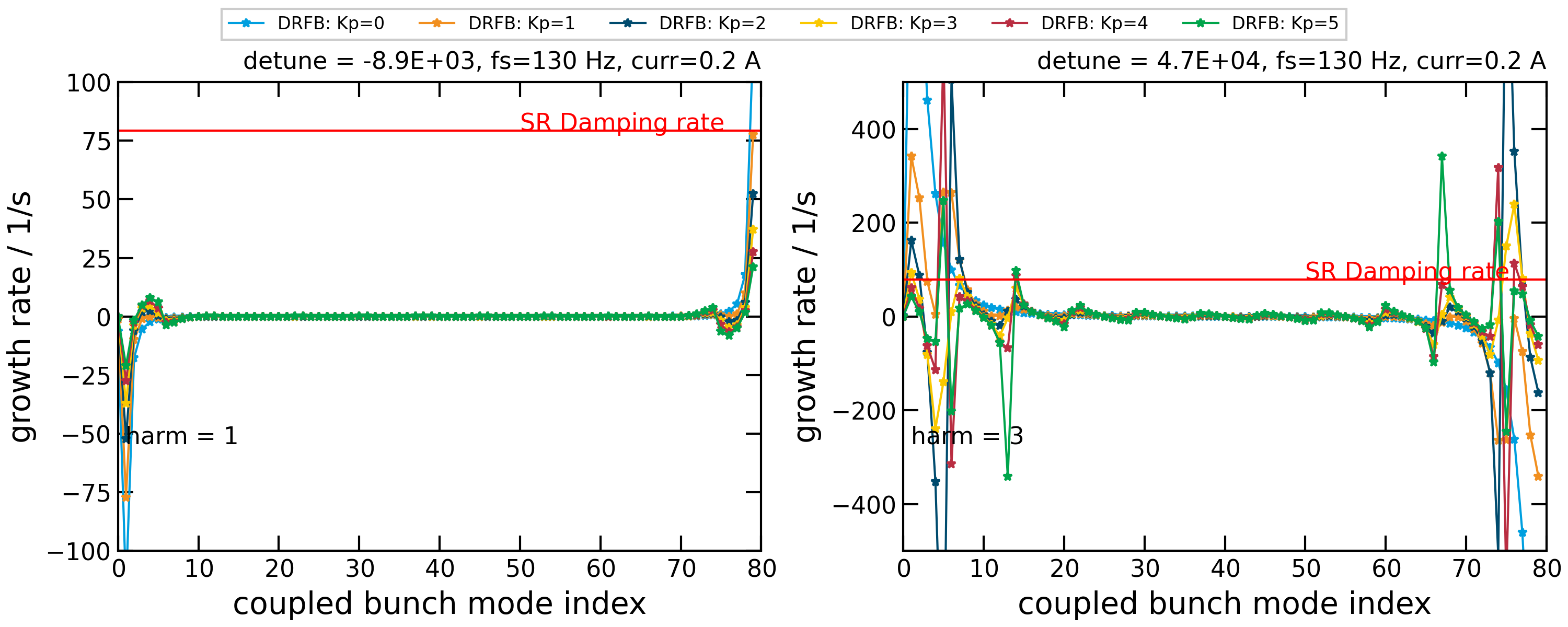}
\caption{\label{fig:5.2}Top: Impedance of the main cavity (left) and the harmonic cavity (right) when the DRF-FB loop is open and closed. The parameters of the PID controller are $K_p=1$, $\tau=1000$ ns in both cavities. Bottom: The CBM growth rate due to the main (left) and the harmonic (right) cavities when the DRF-FB loops are closed with a scan of $K_p$ from 0 to 5. The beam parameters are set the same as those in Fig.~\ref{fig:3.4}. }
\end{figure}

Figure~\ref{fig:5.2} shows the amplitude of the impedance (top) and CBM instability growth rate (bottom) due to the main (left) and the harmonic (right) cavities, when the control loop includes DRF-FB only. The impedance plot on the top is obtained when $K_p=1$ and consists of four groups of curves. The impedance of the open-loop and closed-loop are shown in blue and red. The impedance sampled by the beam at side-bands $f'= p f_0 \pm f_s$ are colored in black and green. As expected, the amplitude of the closed-loop impedance around $f_{rf}$ is reduced by a factor of $K_p+1$. However, additional peaks are created in the closed-loop impedance due to the time delay $\tau=1000$ ns. The amplitudes of these peaks increase when the proportional gain $K_p$ increases, which could further contribute to longitudinal instabilities. The growth rate results include a scan of $K_p$ from 0 to 5 shown in different colors. It indicates that increasing the $K_p$ does not always reduce the effective impedance beam sampled due to the large delay $\tau$. On the contrary, these additional peaks of the closed-loop impedance can excite other CBMs. In this sense, the proportional gain $K_p$ has to be limited to a certain value for the given time delay $\tau$.  With the analytical estimation from Eq.~\ref{eq:4.B.4}, it is found that the impedance reduction factor  $R_{min}/R_{L}\approx $ 0.4 and 2 for the main and harmonic cavities. With $K_p=1$, we observe that the maximum growth rate is reduced to 350 $1/s$.

\subsubsection{OTD-FB}\label{section:5.B.2}
\begin{figure}
\includegraphics[width=1\textwidth]{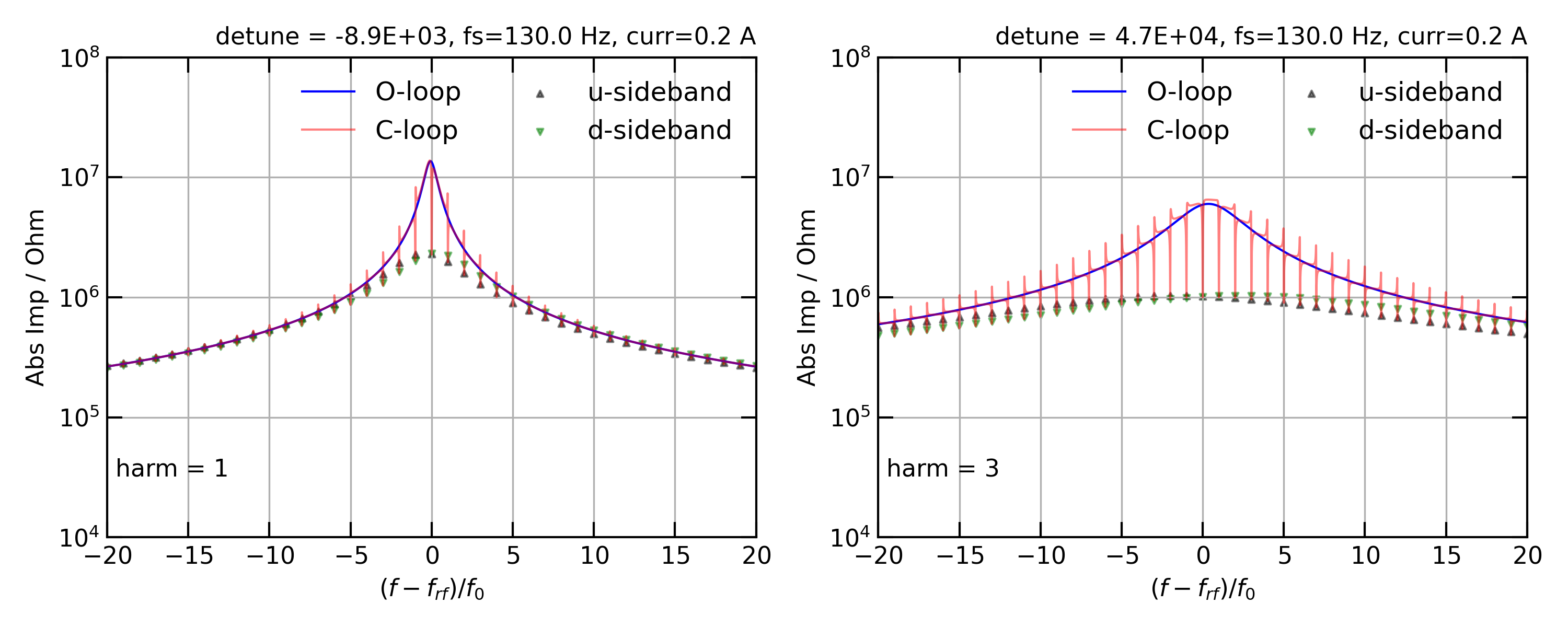}
\includegraphics[width=1\textwidth]{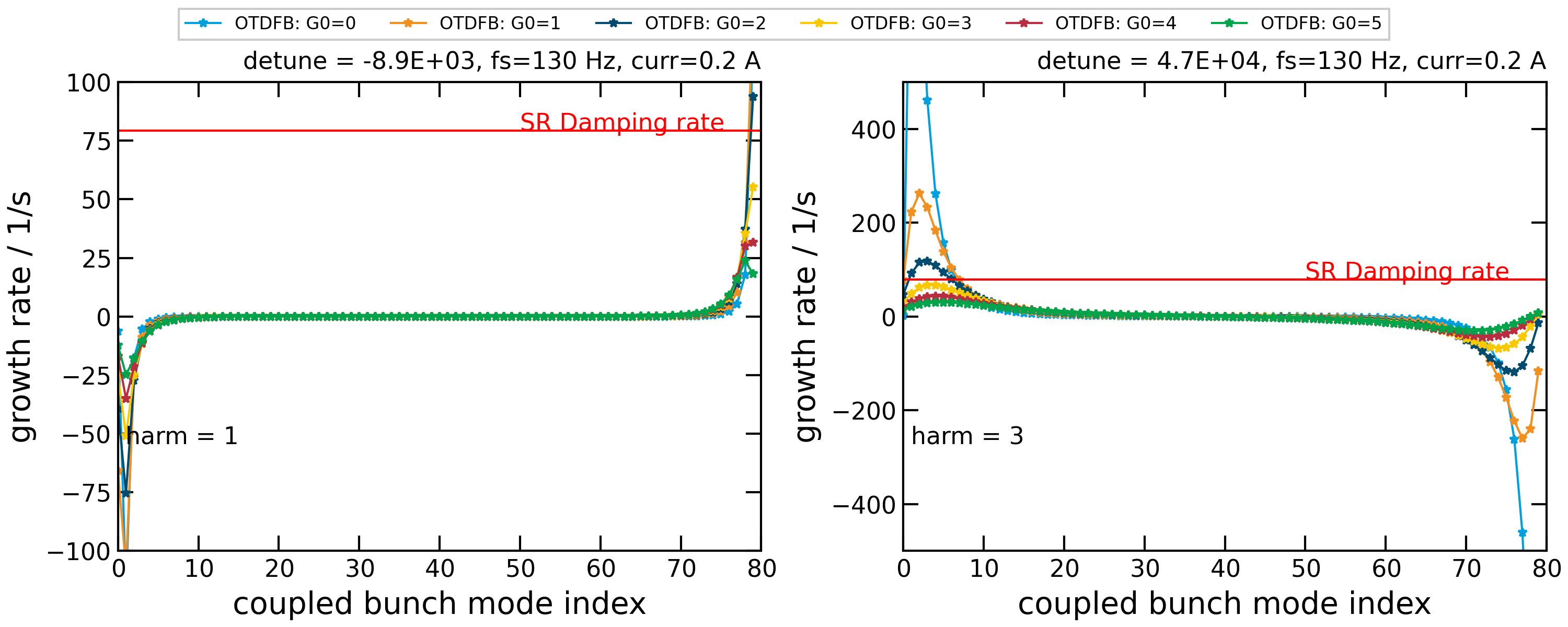}
\caption{\label{fig:5.4}Top: impedance of the main cavity (left) and the harmonic cavity (right) when the OTD-FB loop is open and closed. The parameters of the OTD-FB gain are $G_0=5$, $\alpha=31/32$ in both cavities. Bottom: The CBM growth rate due to the main (left) and the harmonic (right) cavities when the OTD-FB loops are closed with a scan of $G_0$ from 0 to 5. The beam parameters are set to the same as those in Fig.~\ref{fig:3.4}. }
\end{figure}
Similarly, Fig.~\ref{fig:5.4} shows the impedance (top) and  CBM instability growth rate (bottom) due to the main (left) and the harmonic (right) cavities when the control loop includes the OTD-FB only. On the impedance plot, the OTD-FB gain $G_0$ is set to 5. As expected, the closed-loop impedance is roughly reduced by a factor of 6 at every revolution harmonic. The growth rate results include a scan of $G_0$ from 0 to 5 represented by different colors. The growth rate will be further reduced by increasing the OTD-FB gain $G_0$. It indicates that all longitudinal CBM can be stabilized by setting the OTD-FB  gain $G_0\geq5$ in both cavities at 200 mA beam current.

\subsubsection{DRF-FB plus OTD-FB}\label{section:5.B.3}
\begin{figure}
\includegraphics[width=1\textwidth]{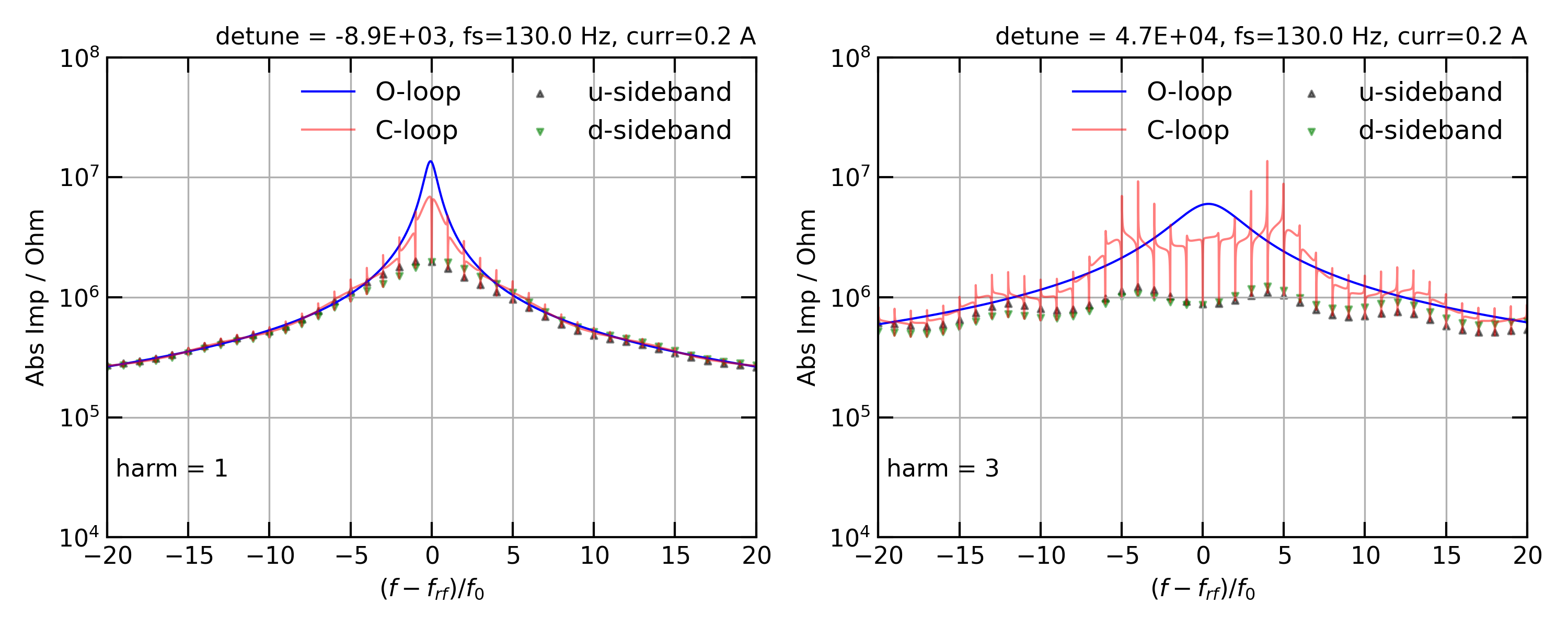}
\includegraphics[width=1\textwidth]{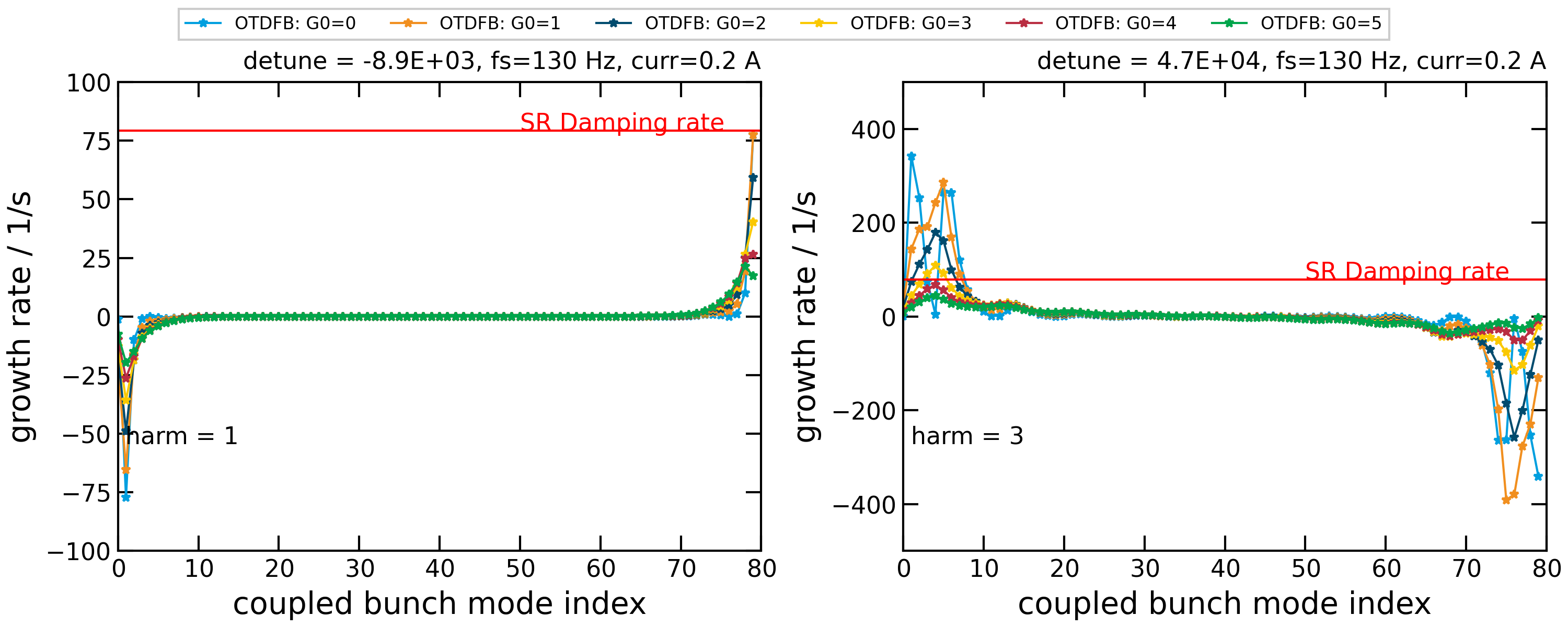}
\caption{\label{fig:5.6}Top: impedance of the main cavity (left) and the harmonic cavity (right) when the parallel LLRF feedback loops are open and closed. The parameters of the OTD-FB gain are $G_0=5$, $\alpha=31/32$, the parameters of the PID controller are $K_p=1$, $\tau=1000$ ns. Bottom: The CBM growth rate due to the main (left) and the harmonic (right) cavities when the LLRF feedback loops are closed with a scan of $G_0$ from 0 to 5. The beam parameters are set the same as those in Fig.~\ref{fig:3.4}. }
\end{figure}

The nominal LLRF feedback design for PETRA-IV includes both the DRF-FB and the OTD-FB, where the feedback loops are combined in parallel, so that the transformation is $H(s)=H_{OTDFB}(s)+H_{DRFB}(s)$. Fig.~\ref{fig:5.6} shows the impedance (top) and CBM instability growth rate (bottom) due to the main (left) and the harmonic (right) cavities. The parameter of the OTD-FB is $G_0=5$, and the parameters of the DRF-FB are $K_p=1$, $\tau=1000$ ns. The growth rate results include a scan of OTD-FB $G_0$ from 0 to 5 shown different colors.  The growth rate will be further reduced by increasing the OTD-FB gain $G_0$. It indicates that all longitudinal CBM can be stabilized by setting the OTD-FB  gain $G_0\geq5$ in both cavities at 200 mA beam current. 

When the feedback loops are combined in series, the total transfer function becomes $H(s)=H_{OTDFB}(s)H_{DRFB}(s)$. If the feedback parameters are carefully chosen such that the total time delay equals one revolution period $T_0$, the transfer function $H(s)$ becomes equivalent to that of the OTD-FB configuration.

\begin{figure}
\includegraphics[width=1\textwidth]{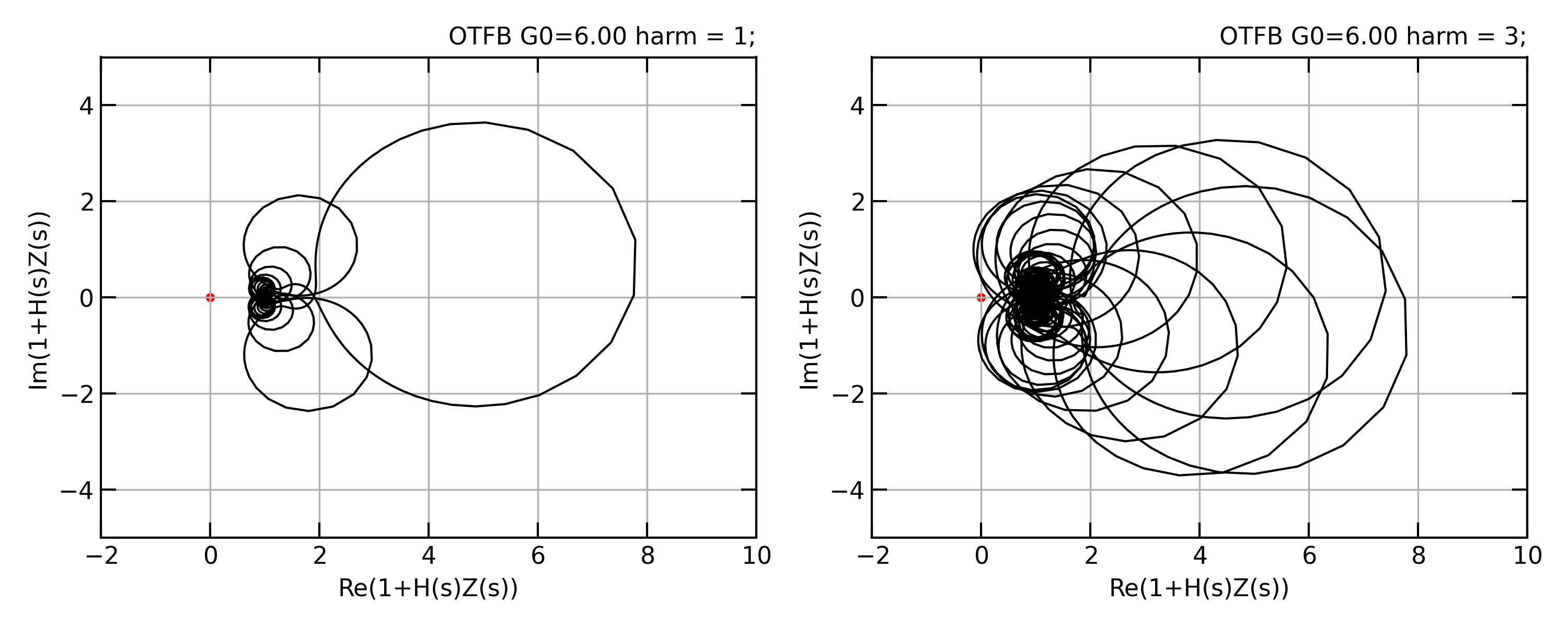}
\caption{\label{fig:5.0} Nyquist plot at the main cavity (left) and the harmonic cavity (right) when the DRF-FB loop is closed. The parameters of the PID controller are $K_p=1$, $\tau=1000$ ns. The parameters of the OTD-FB are $\alpha=31/32$, $G_0=6$, the frequency swept is in the range of $f_{rf}\pm 30 f_0$.}
\end{figure}

Fig.~\ref{fig:5.0} shows the Nyquist plot at the main cavity (left) and the harmonic cavity (right) when the feedback loop is closed, and the DRF-FB and OTD-FB are combined in parallel. The parameters of the PID controller are $K_p=1$, $\tau=1000$ ns. The parameters of the OTD-FB are $\alpha=31/32$, $G_0=6$. The frequency is swept over the range of $f_{rf}\pm 30 f_0$. The feedback loop is stable since the Nyquist diagram does not surround the point (0,0). 

\subsection{Verification of instability mitigation by LLRF feedback in simulation}\label{section:5.C}
\begin{figure}
\includegraphics[width=1\textwidth]{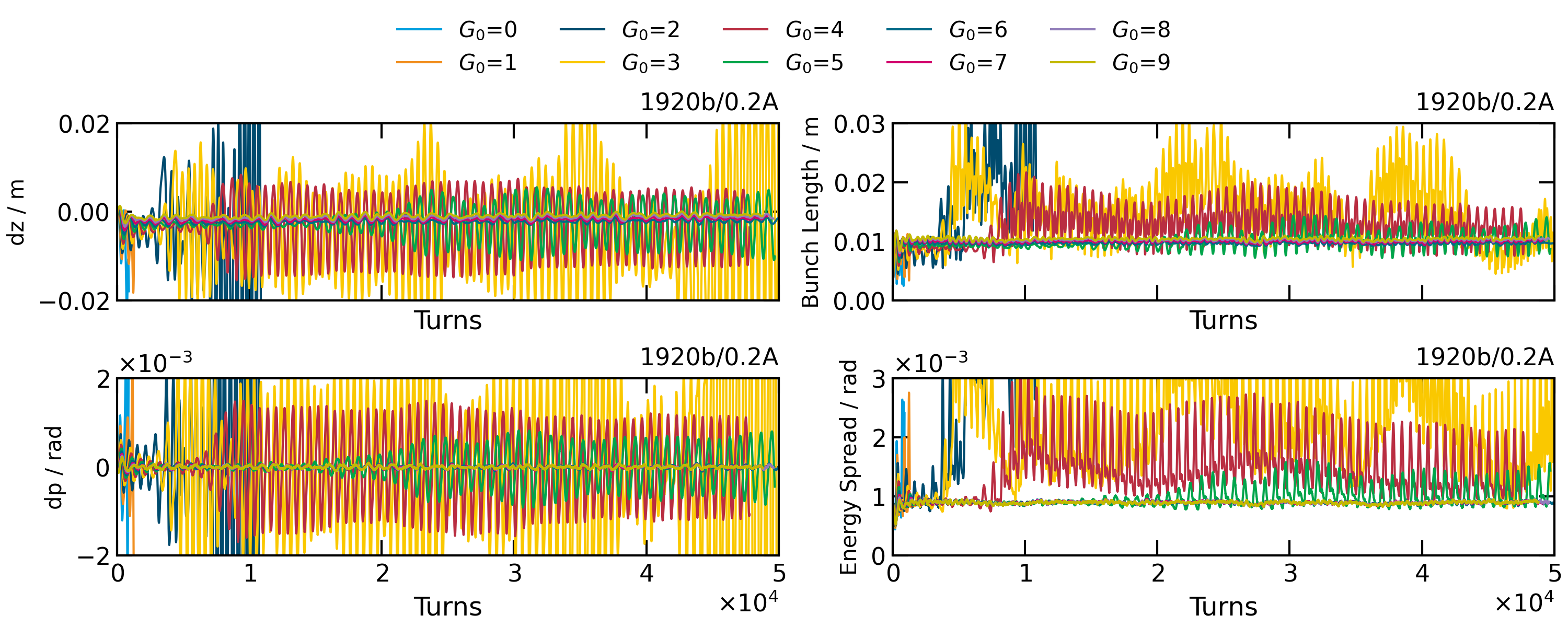}
\caption{\label{fig:5.8} Bunch center (top-left), bunch length (top-right), bunch energy center (bottom-left) and energy spread (bottom-right) as a function of tracking turns. The parameters of the PID controller are $K_p=1$, $\tau=1000$ ns. The parameter of the OTD-FB $\alpha=31/32$. Different lines represent a scan of the OTD-FB gain $G_0$ from 0 to 9. The fill pattern is set as the brightness mode, where 1920 bunches evenly occupy the ring with 200 mA beam current.}
\end{figure}

In the numerical simulations, the parallel LLRF feedback configuration is chosen as an example to demonstrate how longitudinal beam motion can be stabilized by LLRF feedback. For both cavities, we set the DRF-FB parameters to $K_p=1$ and $\tau=1000$~ns, set the OTD-FB parameter to $\alpha=31/32$, and leave the gain $G_0$ as a free parameter. In the simulations, the brightness-mode filling pattern (1920 bunches/200~mA) is used, with 1920 bunches evenly distributed around the ring and a total beam current of 200~mA. Taking the first bunch as an example, Fig.~\ref{fig:5.8} shows its longitudinal centroid (top-left), rms bunch length (top-right), mean energy deviation (bottom-left), and energy spread (bottom-right) as functions of the number of tracking turns. Different colors represent a scan of the OTD-FB gain $G_0$ from 0 to 9. When the OTD-FB gain is $G_0>5$, the beam is stabilized with an equilibrium bunch length of approximately 10~mm, lengthened roughly by a factor of 5 as expected. Meanwhile, the beam remains well centered in both longitudinal position ($dz$) and energy ($dp$). The energy spread remains close to its nominal value. Once the beam is stabilized, we do not observe any significant interference between the different LLRF feedback loops in either the main or harmonic cavity.

However, there is still some discrepancy between the required OTD-FB gain for instability mitigation found in theory and in simulation. It could be due to the filling pattern setting, accuracy in estimation of $f_s$, noise in simulation, $etc$. Considering the above, the simulation results are in reasonable agreement with the analytical predictions  

\begin{figure}
\includegraphics[width=1\textwidth]{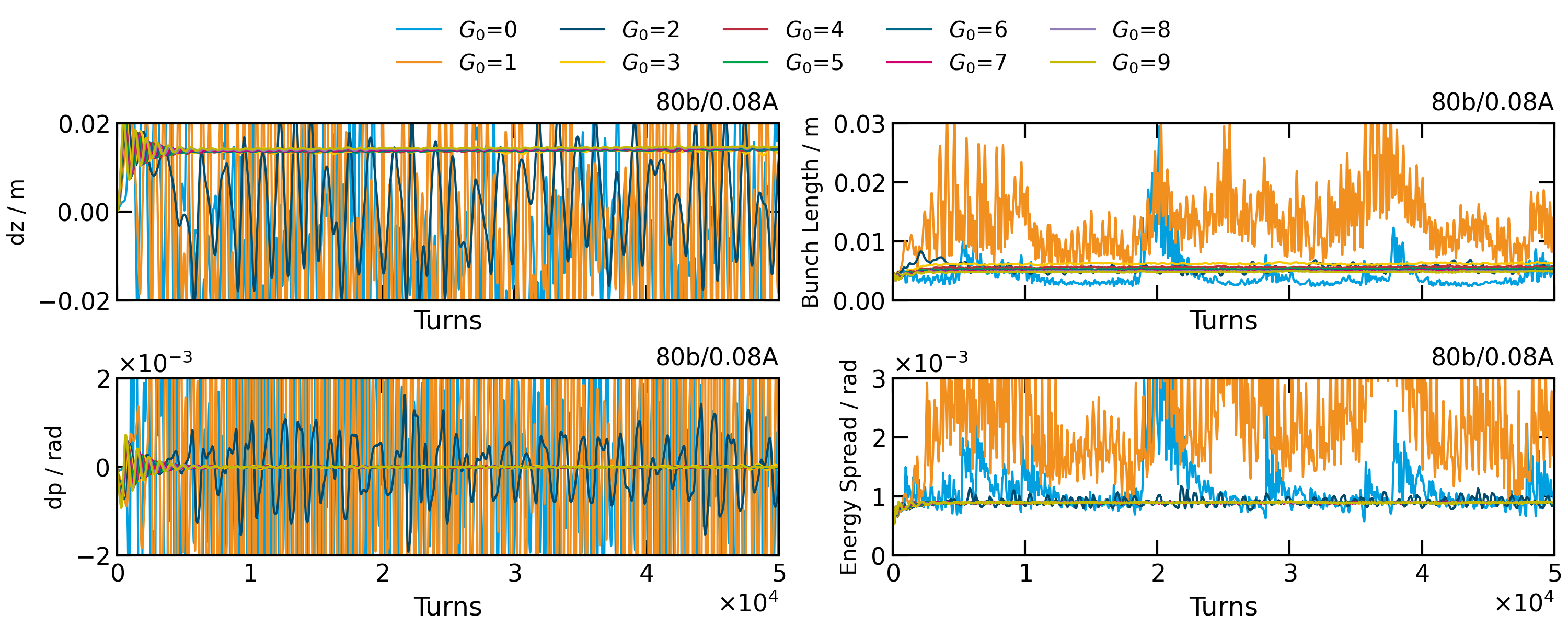}
\caption{\label{fig:5.9} Bunch center (top-left), bunch length (top-right), bunch energy center (bottom-left) and energy spread (bottom-right) as a function of tracking turns. The parameters of the PID controller are $K_p=1$, $\tau=1000$ ns. The parameter of the OTD-FB $\alpha=31/32$. Different lines represent a scan of the OTD-FB gain $G_0$ from 0 to 9. The fill pattern is set as the timing mode, where 80 bunches evenly occupy the ring with 80 mA beam current.}
\end{figure}

With the same LLRF feedback configuration, we also assess the beam performance in the timing operation mode (80 bunches/80~mA). Compared with the brightness mode, the single-bunch charge increases by a factor of 10, while the bunch spacing increases by a factor of 24. Consequently, the variation of the beam-induced voltage from bunch to bunch in the timing mode is more significant, which can degrade the performance of the LLRF feedback loops. Fig.~\ref{fig:5.9} shows the same beam characteristics as Fig.~\ref{fig:5.8}. The beam can be stabilized by setting the OTD-FB gain to $G_0\geq3$. For a stable case, after the beam reaches an equilibrium state following $5\times10^4$ tracking turns, the beam exhibits a longitudinal position offset of $dz\approx15$~mm. Consequently, the cavity voltage $\boldsymbol{\tilde{V}_c}$ experienced by the beam deviates significantly from that required by the ideal bunch-lengthening condition, resulting in a bunch length of approximately 5~mm, corresponding to a bunch-lengthening factor of about 2. To achieve greater bunch lengthening, a finer scan of the OTD-FB gain together with a readjustment of the reference cavity voltage is required.

\begin{figure}
\includegraphics[width=0.48\textwidth]{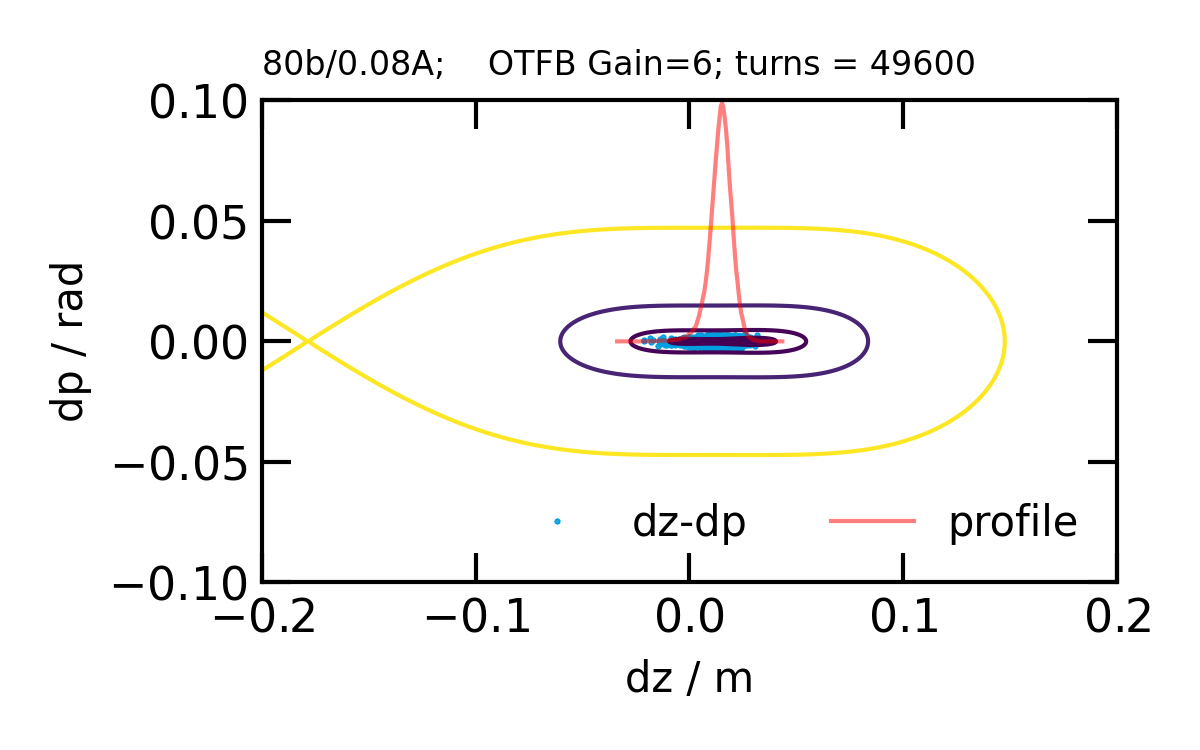}
\includegraphics[width=0.48\textwidth]{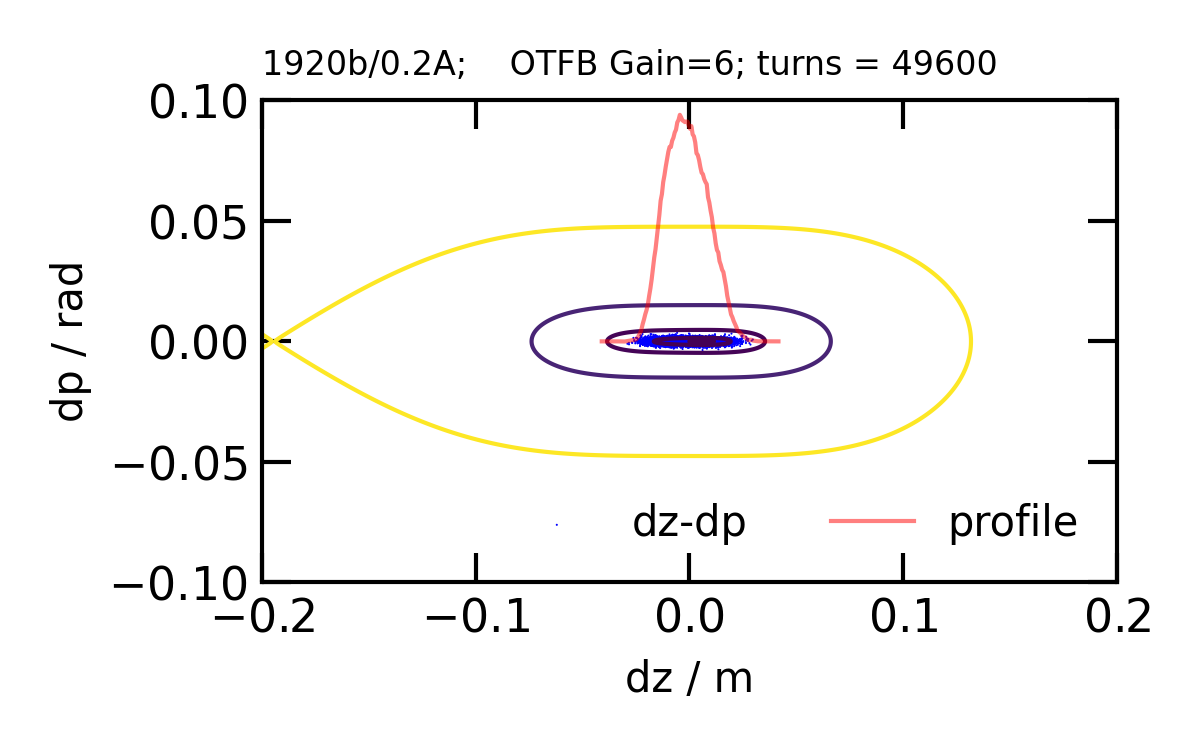}
\caption{\label{fig:5.10}The beam distribution (blue) and Hamiltonian tori (colorful) and beam profile (red) in the longitudinal phase space for timing mode (left) and bright mode (right) at the equilibrium state. The left/right subplot is simulated with the timing/brightness operation mode. The parameters of the PID controller are $K_p=1$, $\tau=1000$ ns; the parameters of the OTD-FB $\alpha=31/32$, $G_0=6$.}
\end{figure}

Fig.~\ref{fig:5.10} shows a comparison of the beam distributions and Hamiltonian tori in longitudinal phase space for the timing mode (left) and brightness mode (right) at equilibrium, with an OTD-FB gain of $G_0=6$.

In practice, the cavities are expected to operate under the optimal detuning condition to minimize the required generator power. The optimal detuning phase is given by
\begin{equation}
\label{eq:5.C.1}
\psi = \arctan(-\frac{V_{br}}{\boldsymbol{|\tilde{V}_c}|} \sin{\phi_s}).  
\end{equation}
Under this condition, the detuning frequencies are $-27/-277$~kHz for the main and harmonic cavities, respectively, resulting in stronger longitudinal CBM instabilities. We also perform numerical simulations with ELEGANT under these optimal detuning conditions. The results show that the beam can again be stabilized when the OTD-FB gain satisfies $G_0\geq10$.

\section{Summary and conclusions}\label{section:6}
The coupled beam–cavity–feedback system exhibits complex dynamics, and careful attention must be paid to the control of longitudinal beam instabilities. In this paper, using the nominal dual-RF system in PETRA-IV as an example, we discuss the coupled longitudinal dynamics in general. The nominal PETRA-IV design includes both main and harmonic cavities, which are normal-conducting and work actively. Due to the large shunt impedance and low quality factor, the harmonic cavity drives unstable longitudinal CBM, which limit the maximum achievable beam current. LLRF feedback is employed to mitigate these coupled-bunch instabilities. For the DRF-FB, the large system delay ($\tau = 1000$ ns) restricts the proportional gain $K_p$ to 1. With DRF-FB only, the beam still suffers from coupled-bunch instabilities. Nevertheless, DRF-FB remains necessary for compensating disturbances from other sources. The OTD-FB is more effective for instability suppression, as it reduces the impedance sampled by the beam at synchrotron sidebands. In this work, we propose applying DRF-FB and OTD-FB in parallel; the longitudinal CBM instability can be mitigated when the OTD-FB gain is set above 5 or 10, according to the cavity configurations. 

This work further emphasizes the essential role of low-level radio-frequency (LLRF) systems in future synchrotron light sources in which active RF systems are employed. As beam-stability requirements become increasingly stringent to achieve higher brilliance and lower emittance, advanced LLRF control will be critical to maintaining longitudinal stability, suppressing coupled-bunch instabilities, and ensuring reliable machine operation. The methodology presented in this study is broadly applicable and can be extended to a wide range of accelerator facilities equipped with active RF systems.

\section{Acknowledgements}
The authors thank R. A. Jameson for proofreading the manuscript and providing numerous helpful comments and suggestions. The authors thank the helpful discussion from Peter Hülsmann from DESY RF group. Thanks to Ivan Karpov from CERN for the helpful discussion on the one-turn feedback. The ongoing collaboration with MHF group was a guiding principle of studying CBM instability caused by the fundamental mode impedance. We appreciate the leadership of R. Bartollini, coordinating the group efforts of MPY, MHF, and MSK within the PETRA IV project. This research was supported by the Maxwell computational resources operated at Deutsches Elektronen-Synchrotron DESY, Hamburg, Germany.

%\section{References}
% \medskip
% \bibliographystyle{plain}
\bibliographystyle{ieeetr}
\bibliography{rfnote}

@misc{elegantWebPage,
  title = {},
  howpublished = {\url{https://ops.aps.anl.gov/manuals/elegant_latest/elegant.html}}
}

@article{boussard1985control,
  title={Control of cavities with high beam loading},
  author={Boussard, Daniel},
  journal={IEEE Transactions on Nuclear Science},
  volume={32},
  number={5},
  pages={1852--1856},
  year={1985},
  publisher={IEEE}
}

@techreport{baudrenghien2001low,
  title={Low level RF systems for synchrotrons: part II: High Intensity. Compensation of the beam induced effects},
  author={Baudrenghien, P},
  year={2001},
  institution={CERN-SL-Note-2001-028-HRF}
}

@mastersthesis{schilcher1998vector,
    author = "Schilcher, T.",
    title = "{Vector sum control of pulsed accelerating fields in Lorentz force detuned superconducting cavities}",
    reportNumber = "DESY-TESLA-98-20",
    type = "Other thesis",
    month = "8",
    year = "1998",
    school = "Hamburg Univeristy"
}

@article{berenc2015modeling,
  title={Modeling RF feedback in Elegant for bunch-lengthening studies for the advanced photon source upgrade},
  author={Berenc, Tim and Borland, Michael and Lindberg, RR},
  journal={Proc. of IPAC15, MOPMA006},
  year={2015}
}

@article{LI2025170031,
title = {CETASim: A numerical tool for beam collective effect study in storage rings},
journal = {NIMA},
volume = {1070},
pages = {170031},
year = {2025},
issn = {0168-9002},
doi = {https://doi.org/10.1016/j.nima.2024.170031},
url = {https://www.sciencedirect.com/science/article/pii/S0168900224009574},
author = {Chao Li and Yong-Chul Chae},
}

@misc{CETASim,
  title = {},
  howpublished = {\url{https://github.com/ESR-BeamSimulation/CETASIM}}
}

@article{byrd1995spectral,
  title={Spectral characterization of longitudinal coupled-bunch instabilities at the Advanced Light Source},
  author={Byrd, JM and Corlett, J},
  journal={Part. Accel.},
  volume={51},
  pages={29--42},
  year={1995}
}

@book{AlexChao,
  title={Physics of collective beam instabilities in high energy accelerators},
  author={Chao, Alexander Wu},
  year={1993},
  publisher={Wiley-Interscience Publication}  
}

@book{ng2006physics,
  title={Physics of intensity dependent beam instabilities},
  author={Ng, King-Yuen},
  year={2006},
  publisher={World Scientific}
}

@article{PhysRevSTAB.17.064401,
  title = {Equilibrium bunch density distribution with passive harmonic cavities in a storage ring},
  author = {Tavares, Pedro F. and Andersson, \AA{}ke and Hansson, Anders and Breunlin, Jonas},
  journal = {Phys. Rev. ST Accel. Beams},
  volume = {17},
  issue = {6},
  pages = {064401},
  numpages = {14},
  year = {2014},
  month = {Jun},
  publisher = {American Physical Society},
  doi = {10.1103/PhysRevSTAB.17.064401},
  url = {https://link.aps.org/doi/10.1103/PhysRevSTAB.17.064401}
}

@article{yamaguchi2023systematic,
  title={Systematic study on the static Robinson instability in an electron storage ring},
  author={Yamaguchi, Takaaki and Sakanaka, Shogo and Yamamoto, Naoto and Naito, Daichi and Takahashi, Takeshi},
  journal={Physical Review Accelerators and Beams},
  volume={26},
  number={4},
  pages={044401},
  year={2023},
  publisher={APS}
}

@article{MIYAHARA1987518,
title = {Equilibrium phase instability in the double RF system for Landau damping},
journal = {Nuclear Instruments and Methods in Physics Research Section A: Accelerators, Spectrometers, Detectors and Associated Equipment},
volume = {260},
number = {2},
pages = {518-528},
year = {1987},
issn = {0168-9002},
doi = {https://doi.org/10.1016/0168-9002(87)90125-2},
url = {https://www.sciencedirect.com/science/article/pii/0168900287901252},
author = {Y. Miyahara and S. Asaoka and A. Mikuni and K. Soda},
}

@article{PhysRevAccelBeams.27.044403,
  title = {Experimental observation of a mode-1 instability driven by Landau cavities in a storage ring},
  author = {Cullinan, F. J. and Andersson, \AA{}. and Breunlin, J. and Brosi, M. and Tavares, P. F.},
  journal = {Phys. Rev. Accel. Beams},
  volume = {27},
  issue = {4},
  pages = {044403},
  numpages = {13},
  year = {2024},
  month = {Apr},
  publisher = {American Physical Society},
  doi = {10.1103/PhysRevAccelBeams.27.044403},
  url = {https://link.aps.org/doi/10.1103/PhysRevAccelBeams.27.044403}
}

@article{Haissinski,
  title = {Exact Longitudinal Equilibrium Distribution of Stored Electrons in the Presence of Self-Fields},
  author = {J. Haissinski},
  journal = {Nuovo Cimento 18B},
  volume = {72},
  year = {1973},
 }

@article{Pedersen:244817,
      author        = "Pedersen, F",
      title         = "{rf cavity feedback}",
      journal       = "CERN-PS-92-59-RF",
      year          = "1992",
      url           = "https://cds.cern.ch/record/244817",
}

@misc{Z-Transform,
  title = {},
  howpublished = {\url{https://en.wikipedia.org/wiki/Z-transform}}
}

@article{PhysRevSTAB.4.074401,
  title = {Robinson instabilities with a higher-harmonic cavity},
  author = {Bosch, R. A. and Kleman, K. J. and Bisognano, J. J.},
  journal = {Phys. Rev. ST Accel. Beams},
  volume = {4},
  issue = {7},
  pages = {074401},
  numpages = {23},
  year = {2001},
  month = {Jul},
  publisher = {American Physical Society},
  doi = {10.1103/PhysRevSTAB.4.074401},
  url = {https://link.aps.org/doi/10.1103/PhysRevSTAB.4.074401}
}

@article{PhysRevAccelBeams.27.064402,
  title = {Analytic formulas for the $D$-mode Robinson instability},
  author = {He, Tianlong and Li, Weiwei and Bai, Zhenghe and Li, Weimin},
  journal = {Phys. Rev. Accel. Beams},
  volume = {27},
  issue = {6},
  pages = {064402},
  numpages = {10},
  year = {2024},
  month = {Jun},
  publisher = {American Physical Society},
  doi = {10.1103/PhysRevAccelBeams.27.064402},
  url = {https://link.aps.org/doi/10.1103/PhysRevAccelBeams.27.064402}
}

@article{alves2025theoretical,
  title={Theoretical models for longitudinal coupled-bunch instabilities driven by harmonic cavities in electron storage rings},
  author={Alves, Murilo B},
  journal={Physical Review Accelerators and Beams},
  volume={28},
  number={3},
  pages={034401},
  year={2025},
  publisher={APS}
}

@article{agapov2024beam,
  title={Beam dynamics performance of the proposed PETRA IV storage ring},
  author={Agapov, I and Antipov, S and Bartolini, R and Brinkmann, R and Chae, YC and Cortes-Garcia, EC and Einfeld, D and Hellert, T and Huening, M and Jebramcik, MA and others},
  journal={arXiv preprint arXiv:2408.07995},
  year={2024}
}

@article{PhysRevSTAB.12.120701,
  title = {Beam dynamics and expected performance of Sweden's new storage-ring light source: MAX IV},
  author = {Leemann, S. C. and Andersson, \AA{}. and Eriksson, M. and Lindgren, L.-J. and Wall\'en, E. and Bengtsson, J. and Streun, A.},
  journal = {Phys. Rev. ST Accel. Beams},
  volume = {12},
  issue = {12},
  pages = {120701},
  numpages = {15},
  year = {2009},
  month = {Dec},
  publisher = {American Physical Society},
  doi = {10.1103/PhysRevSTAB.12.120701},
  url = {https://link.aps.org/doi/10.1103/PhysRevSTAB.12.120701}
}

@article{PhysRevAccelBeams.24.110701,
  title = {Commissioning of the hybrid multibend achromat lattice at the European Synchrotron Radiation Facility},
  author = {Raimondi, P. and Carmignani, N. and Carver, L. R. and Chavanne, J. and Farvacque, L. and Le Bec, G. and Martin, D. and Liuzzo, S. M. and Perron, T. and White, S.},
  journal = {Phys. Rev. Accel. Beams},
  volume = {24},
  issue = {11},
  pages = {110701},
  numpages = {15},
  year = {2021},
  month = {Nov},
  publisher = {American Physical Society},
  doi = {10.1103/PhysRevAccelBeams.24.110701},
  url = {https://link.aps.org/doi/10.1103/PhysRevAccelBeams.24.110701}
}

@article{sajaev2025commissioning,
  title={Commissioning of the Advanced Photon Source upgrade-the first swap-out injection-based synchrotron light source},
  author={Sajaev, Vadim and Borland, M and Calvey, J and Dooling, J and Emery, L and Harkay, K and Kuklev, N and Mohsen, O and Sun, Y and Arnold, N and others},
  journal={Proc. IPAC’25},
  pages={915--918},
  year={2025}
}

@article{boge2025sls,
  title={{SLS} 2.0 storage ring commissioning},
  author={B{\"o}ge, Michael and others},
  journal={Proc. IPAC’25},
  pages={1714--1717},
  year={2025}
}

@article{PhysRevAccelBeams.24.011002,
  title = {Thresholds for loss of Landau damping in longitudinal plane},
  author = {Karpov, Ivan and Argyropoulos, Theodoros and Shaposhnikova, Elena},
  journal = {Phys. Rev. Accel. Beams},
  volume = {24},
  issue = {1},
  pages = {011002},
  numpages = {19},
  year = {2021},
  month = {Jan},
  publisher = {American Physical Society},
  doi = {10.1103/PhysRevAccelBeams.24.011002},
  url = {https://link.aps.org/doi/10.1103/PhysRevAccelBeams.24.011002}
}

@article{herr2016introduction,
  title={Introduction to Landau damping},
  author={Herr, Werner},
  journal={arXiv preprint arXiv:1601.05227},
  year={2016}
}

@article{landau196561,
  title={61--On the vibrations of the electronic plasma},
  author={Landau, L and Ter Haar, D},
  journal={The Collected Papers of LD Landau},
  pages={445--460},
  year={1965},
  publisher={Pergamon,}
}

@article{PhysRevAccelBeams.21.114404,
  title = {Passive higher-harmonic rf cavities with general settings and multibunch instabilities in electron storage rings},
  author = {Venturini, Marco},
  journal = {Phys. Rev. Accel. Beams},
  volume = {21},
  issue = {11},
  pages = {114404},
  numpages = {14},
  year = {2018},
  month = {Nov},
  publisher = {American Physical Society},
  doi = {10.1103/PhysRevAccelBeams.21.114404},
  url = {https://link.aps.org/doi/10.1103/PhysRevAccelBeams.21.114404}
}

@article{lindberg2018theory,
  title={Theory of coupled-bunch longitudinal instabilities in a storage ring for arbitrary rf potentials},
  author={Lindberg, Ryan R},
  journal={Physical Review Accelerators and Beams},
  volume={21},
  number={12},
  pages={124402},
  year={2018},
  publisher={APS}
}

@article{PEREZ2025170195,
title = {Active harmonic EU cavity: Commissioning and operation with beam},
journal = {Nuclear Instruments and Methods in Physics Research Section A: Accelerators, Spectrometers, Detectors and Associated Equipment},
volume = {1072},
pages = {170195},
year = {2025},
issn = {0168-9002},
doi = {https://doi.org/10.1016/j.nima.2024.170195},
url = {https://www.sciencedirect.com/science/article/pii/S0168900224011215},
author = {F. Perez and J. Alvarez and I. Bellafont and B. Bravo and J. Ocampo and A. Salom and P. Solans and M. Ebert and N.-O. Fröhlich and P. Hülsmann and R. Onken and W. Anders and V. Duerr and T. Löwner and A. Matveenko and M. Ries and L. Shi and Y. Tamashevich and A. Tsakanian and H. {De Gersem} and W. Müller}
}

@article{PhysRevAccelBeams.28.054401,
  title = {Semianalytical algorithms to study longitudinal beam instabilities in double rf systems},
  author = {Gamelin, A. and Gubaidulin, V. and Alves, M. B. and Olsson, T.},
  journal = {Phys. Rev. Accel. Beams},
  volume = {28},
  issue = {5},
  pages = {054401},
  numpages = {21},
  year = {2025},
  month = {May},
  publisher = {American Physical Society},
  doi = {10.1103/PhysRevAccelBeams.28.054401},
  url = {https://link.aps.org/doi/10.1103/PhysRevAccelBeams.28.054401}
}

@article{PhysRevAccelBeams.20.011004,
  title = {Fundamental cavity impedance and longitudinal coupled-bunch instabilities at the High Luminosity Large Hadron Collider},
  author = {Baudrenghien, P. and Mastoridis, T.},
  journal = {Phys. Rev. Accel. Beams},
  volume = {20},
  issue = {1},
  pages = {011004},
  numpages = {13},
  year = {2017},
  month = {Jan},
  publisher = {American Physical Society},
  doi = {10.1103/PhysRevAccelBeams.20.011004},
  url = {https://link.aps.org/doi/10.1103/PhysRevAccelBeams.20.011004}
}

@article{PhysRevSTAB.13.052802,
  title = {Lessons learned from positron-electron project low level rf and longitudinal feedback},
  author = {Fox, J. and Mastorides, T. and Rivetta, C. and Van Winkle, D. and Teytelman, D.},
  journal = {Phys. Rev. ST Accel. Beams},
  volume = {13},
  issue = {5},
  pages = {052802},
  numpages = {16},
  year = {2010},
  month = {May},
  publisher = {American Physical Society},
  doi = {10.1103/PhysRevSTAB.13.052802},
  url = {https://link.aps.org/doi/10.1103/PhysRevSTAB.13.052802}
}

@article{PhysRevAccelBeams.26.114602,
  title = {Beam longitudinal dynamics simulation studies},
  author = {Timko, H. and Albright, S. and Argyropoulos, T. and Damerau, H. and Iliakis, K. and Intelisano, L. and Karlsen-Baeck, B. E. and Karpov, I. and Lasheen, A. and Medina, L. and Quartullo, D. and Repond, J. and Vanel, A. L. and M\"uller, J. Esteban and Schwarz, M. and Tsapatsaris, P. and Typaldos, G.},
  journal = {Phys. Rev. Accel. Beams},
  volume = {26},
  issue = {11},
  pages = {114602},
  numpages = {20},
  year = {2023},
  month = {Nov},
  publisher = {American Physical Society},
  doi = {10.1103/PhysRevAccelBeams.26.114602},
  url = {https://link.aps.org/doi/10.1103/PhysRevAccelBeams.26.114602}
}

@article{PhysRevAccelBeams.23.074402,
  title = {Harmonic-cavity stabilization of longitudinal coupled-bunch instabilities with a nonuniform fill},
  author = {Cullinan, F. J. and Andersson, \AA{}. and Tavares, P. F.},
  journal = {Phys. Rev. Accel. Beams},
  volume = {23},
  issue = {7},
  pages = {074402},
  numpages = {14},
  year = {2020},
  month = {Jul},
  publisher = {American Physical Society},
  doi = {10.1103/PhysRevAccelBeams.23.074402},
  url = {https://link.aps.org/doi/10.1103/PhysRevAccelBeams.23.074402}
}

@article{karpov2023longitudinal,
  title={Longitudinal mode-coupling instabilities of proton bunches in the CERN Super Proton Synchrotron},
  author={Karpov, Ivan},
  journal={Physical Review Accelerators and Beams},
  volume={26},
  number={1},
  pages={014401},
  year={2023},
  publisher={APS}
}

@article{chin1982instability,
  title={Instability of a bunched beam with synchrotron frequency spread},
  author={Chin, Yongho and Yokoya, K and Satoh, K},
  journal={Part. Accel.},
  volume={13},
  number={KEK-82-18},
  pages={45--66},
  year={1982}
}

@inproceedings{wang1980condition,
  title={On the condition for a single bunch high frequency fast blow-up},
  author={Wang, JM and Pellegrini, C},
  booktitle={11th International Conference on High-Energy Accelerators: Geneva, Switzerland, July 7--11, 1980},
  pages={554--561},
  year={1980},
  organization={Springer}
}

@article{krinsky1983longitudinal,
  title={Longitudinal instabilities with a non-harmonic rf potential},
  author={Krinsky, S and Wang, JM},
  journal={IEEE Transactions on Nuclear Science},
  volume={30},
  number={4},
  pages={2492--2494},
  year={1983},
  publisher={IEEE}
}

@techreport{oide1990longitudinal,
  title={Longitudinal single-bunch instability in electron storage rings},
  author={Oide, Katsunobu and Yokoya, Kaoru},
  year={1990}
}

@article{fox2010lessons,
  title={Lessons learned from positron-electron project low level rf and longitudinal feedback},
  author={Fox, J and Mastorides, Themis and Rivetta, C and Van Winkle, D and Teytelman, D},
  journal={Physical Review Special Topics—Accelerators and Beams},
  volume={13},
  number={5},
  pages={052802},
  year={2010},
  publisher={APS}
}

@article{drago2003longitudinal,
  title={Longitudinal quadrupole instability and control in the Frascati DA $\Phi$ NE electron ring},
  author={Drago, A and Gallo, A and Ghigo, A and Zobov, M and Fox, JD and Teytelman, D},
  journal={Physical Review Special Topics—Accelerators and Beams},
  volume={6},
  number={5},
  pages={052801},
  year={2003},
  publisher={APS}
}

@inproceedings{wang1989robinson,
  title={Robinson instability and longitudinal mode coupling},
  author={Wang, T-SF},
  booktitle={Proceedings of the 1989 IEEE Particle Accelerator Conference,.'Accelerator Science and Technology},
  pages={1160--1162},
  year={1989},
  organization={IEEE}
}

@article{kallakuri2022coupled,
  title={Coupled bunch mode zero correction using orbit measurements and rf system phase feedback},
  author={Kallakuri, P and Brill, A and Carwardine, J and Sereno, N},
  journal={Physical Review Accelerators and Beams},
  volume={25},
  number={8},
  pages={082801},
  year={2022},
  publisher={APS}
}

@inproceedings{wang2023rf,
  title={RF feedback simulation for Diamond-II using ELEGANT},
  author={Wang, Siwei and Christou, C and Fielder, R and Gu, P and Martin, IPS},
  booktitle={Proc. 14th International Particle Accelerator Conference (IPAC’23)},
  year={2023}
}

@techreport{baudrenghien2012lhc,
  title={The LHC One-Turn Feedback},
  author={Baudrenghien, P and Mastoridis, T and Molendijk, J},
  year={2012}
}
% \printbibliography
\end{document}